\documentclass[12pt,a4paper]{article}
\usepackage{epsfig}
\usepackage{mathrsfs}

\def\hybrid{\topmargin -20pt    \oddsidemargin 0pt
        \headheight 0pt \headsep 0pt
        \textwidth 6.25in       
        \textheight 9.5in       
        \marginparwidth .875in
        \parskip 5pt plus 1pt   \jot = 1.5ex}

\hybrid

\def\baselinestretch{1.2}

\catcode`\@=11

\def\marginnote#1{}
\newcount\hour
\newcount\minute
\newtoks\amorpm
\hour=\time\divide\hour by60
\minute=\time{\multiply\hour by60 \global\advance\minute by-\hour}
\edef\standardtime{{\ifnum\hour<12 \global\amorpm={am}%
        \else\global\amorpm={pm}\advance\hour by-12 \fi
        \ifnum\hour=0 \hour=12 \fi
        \number\hour:\ifnum\minute<10 0\fi\number\minute\the\amorpm}}
\edef\militarytime{\number\hour:\ifnum\minute<10 0\fi\number\minute}

\def\draftlabel#1{{\@bsphack\if@filesw {\let\thepage\relax
   \xdef\@gtempa{\write\@auxout{\string
      \newlabel{#1}{{\@currentlabel}{\thepage}}}}}\@gtempa
   \if@nobreak \ifvmode\nobreak\fi\fi\fi\@esphack}
        \gdef\@eqnlabel{#1}}
\def\@eqnlabel{}
\def\@vacuum{}
\def\draftmarginnote#1{\marginpar{\raggedright\scriptsize\tt#1}}

\def\draft{\oddsidemargin -.5truein
        \def\@oddfoot{\sl preliminary draft \hfil
        \rm\thepage\hfil\sl\today\quad\militarytime}
        \let\@evenfoot\@oddfoot \overfullrule 3pt
        \let\label=\draftlabel
        \let\marginnote=\draftmarginnote
   \def\@eqnnum{(\theequation)\rlap{\kern\marginparsep\tt\@eqnlabel}%
\global\let\@eqnlabel\@vacuum}  }

\def\preprint{\twocolumn\sloppy\flushbottom\parindent 2em
        \leftmargini 2em\leftmarginv .5em\leftmarginvi .5em
        \oddsidemargin -.5in    \evensidemargin -.5in
        \columnsep .4in \footheight 0pt
        \textwidth 10.in        \topmargin  -.4in
        \headheight 12pt \topskip .4in
        \textheight 6.9in \footskip 0pt
        \def\@oddhead{\thepage\hfil\addtocounter{page}{1}\thepage}
        \let\@evenhead\@oddhead \def\@oddfoot{} \def\@evenfoot{} }

\def\numberbysection{\@addtoreset{equation}{section}
        \def\theequation{\thesection.\arabic{equation}}}

\def\underline#1{\relax\ifmmode\@@underline#1\else
        $\@@underline{\hbox{#1}}$\relax\fi}

\def\titlepage{\@restonecolfalse\if@twocolumn\@restonecoltrue\onecolumn
     \else \newpage \fi \thispagestyle{empty}\c@page\z@
        \def\thefootnote{\fnsymbol{footnote}} }

\def\endtitlepage{\if@restonecol\twocolumn \else \newpage \fi
        \def\thefootnote{\arabic{footnote}}
        \setcounter{footnote}{0}}  

\catcode`@=12
\relax

\def\figcap{\section*{Figure Captions\markboth
        {FIGURECAPTIONS}{FIGURECAPTIONS}}\list
        {Figure \arabic{enumi}:\hfill}{\settowidth\labelwidth{Figure
999:}
        \leftmargin\labelwidth
        \advance\leftmargin\labelsep\usecounter{enumi}}}
 \relax
\def\tablecap{\section*{Table Captions\markboth
        {TABLECAPTIONS}{TABLECAPTIONS}}\list
        {Table \arabic{enumi}:\hfill}{\settowidth\labelwidth{Table
999:}
        \leftmargin\labelwidth
        \advance\leftmargin\labelsep\usecounter{enumi}}}
 \relax
\def\reflist{\section*{References\markboth
        {REFLIST}{REFLIST}}\list
        {[\arabic{enumi}]\hfill}{\settowidth\labelwidth{[999]}
        \leftmargin\labelwidth
        \advance\leftmargin\labelsep\usecounter{enumi}}}
 \relax
\makeatletter
\newcounter{pubctr}
\def\publist{\@ifnextchar[{\@publist}{\@@publist}}
\def\@publist[#1]{\list
        {[\arabic{pubctr}]\hfill}{\settowidth\labelwidth{[999]}
        \leftmargin\labelwidth
        \advance\leftmargin\labelsep
        \@nmbrlisttrue\def\@listctr{pubctr}
        \setcounter{pubctr}{#1}\addtocounter{pubctr}{-1}}}
\def\@@publist{\list
        {[\arabic{pubctr}]\hfill}{\settowidth\labelwidth{[999]}
        \leftmargin\labelwidth
        \advance\leftmargin\labelsep
        \@nmbrlisttrue\def\@listctr{pubctr}}}
 \relax
\makeatother
\newskip\humongous \humongous=0pt plus 1000pt minus 1000pt

\newif\ifdtup

\relax

\def\be{\begin{equation}}
\def\ee{\end{equation}}
\def\ba{\begin{eqnarray}}
\def\ea{\end{eqnarray}}

\def\no{\noindent}

\def\IR{\relax{\rm I\kern-.18em R}}

\begin{document}

\renewcommand{\theequation}{\thesection.\arabic{equation}}

\newcommand{\beq}{\begin{equation}}
\newcommand{\eeq}[1]{\label{#1}\end{equation}}
\newcommand{\ber}{\begin{eqnarray}}
\newcommand{\eer}[1]{\label{#1}\end{eqnarray}}
\newcommand{\eqn}[1]{(\ref{#1})}
\begin{titlepage}
\begin{center}

\hfill September 2008\\

\vskip .4in

{\large \bf Energy-momentum/Cotton tensor duality
for $AdS_4$ black holes}

\vskip 0.6in

{\bf Ioannis Bakas}
\vskip 0.2in
{\em Department of Physics, University of Patras \\
GR-26500 Patras, Greece\\
\footnotesize{\tt bakas@ajax.physics.upatras.gr}}\\

\end{center}

\vskip .8in

\centerline{\bf Abstract}

\no
We consider the theory of gravitational quasi-normal modes for general linear
perturbations of $AdS_4$ black holes. Special emphasis is placed on
the effective Schr\"odinger problems for axial and polar perturbations
that realize supersymmetric partner potential barriers on the half-line. Using
the holographic renormalization method, we compute the energy-momentum tensor
for perturbations satisfying arbitrary boundary conditions at spatial infinity
and discuss some aspects of the problem in the hydrodynamic representation.
It is also observed in this general framework that the energy-momentum tensor
of black hole perturbations and the energy momentum tensor of the gravitational
Chern-Simons action (known as Cotton tensor) exhibit an {\em axial-polar
duality} with respect to appropriately chosen supersymmetric partner boundary
conditions on the effective Schr\"odinger wave-functions. This correspondence
applies to perturbations of very large $AdS_4$ black holes with shear viscosity
to entropy density ratio equal to $1/ 4 \pi$, thus providing a dual graviton
description of their hydrodynamic modes. We also entertain the idea that the
purely dissipative modes of black hole hydrodynamics may admit Ricci flow
description in the non-linear regime.

\vfill
\end{titlepage}
\eject

\def\baselinestretch{1.2}
\baselineskip 16 pt
\noindent

\tableofcontents

\newpage

\section{Introduction}
\setcounter{equation}{0}

The question of stability of the Schwarzschild metric against small
perturbations of the geometry arose more than half century ago in the
seminal work of Regge and Wheeler, \cite{wheeler}. Since then, the subject has
grown enormously (see, for instance, the selected works \cite{zerilli},
\cite{vish}, \cite{chandra1}, among many other important contributions) and
developed into what has become known as the theory of quasi-normal modes
(for reviews of the mathematical and physical aspects of the problem see, for
instance, references \cite{chandra2} and \cite{kokko}, respectively).
In recent years, the theory of quasi-normal modes has also been extended to
black-hole solutions in space-times with cosmological constant $\Lambda$, and in
particular to the $AdS_4$ Schwarzschild background, \cite{lemos}, \cite{moss},
which is the subject of this work.

Although the theory of gravitational perturbations of black holes can be studied
systematically in higher dimensions as well, it is important to realize that four
space-time dimensions are rather special in this framework, since they exhibit a
remarkable duality among the two distinct classes of perturbations. The duality
exists irrespective of $\Lambda$ and connects the effective Schr\"odinger problems
that describe the axial and polar perturbations of the metric.
This relation was first discovered
more that thirty years ago, \cite{chandra1} (but see also reference \cite{chandra2}
for an extensive presentation), by considering gravitational perturbations
of the Schwarzschild metric (without cosmological constant) and it gave rise to what
has become known later in the literature as partner potentials in supersymmetric
quantum mechanics, \cite{ed3}, \cite{susy}. The axial-polar relation persists in the
presence of cosmological constant, \cite{lemos}, \cite{moss}, although
supersymmetric quantum mechanics does not necessarily respect the boundary conditions
imposed on the effective wave-functions at spatial infinity. Yet, there is no
fundamental explanation of this occurrence, to the best of our knowledge, and
any new insight into the problem is certainly welcome. Furthermore, there could be
reformulations and/or different manifestations of this duality in areas where
the general theory of quasi-normal modes is applicable.

AdS/CFT correspondence, \cite{malda}, \cite{igor}, \cite{ed1}, and in particular its
generalization to finite temperature field theory, \cite{ed2}, provide such a framework
using the $AdS$ Schwarzschild solution as background geometry on the bulk. In fact,
many well known facts about the thermodynamics of $AdS$ black holes, as they were
originally formulated by Hawking and Page, \cite{don}, found a natural manifestation
in AdS/CFT correspondence, \cite{ed2}. It was subsequently realized  that the theory
of quasi-normal modes also had a natural place in this framework, as it describes small
deviations from the equilibrium state in finite temperature field theory, \cite{gary};
the inverse time scale for return to equilibrium is given by the (minus) imaginary
part of the corresponding quasi-normal mode. Although scalar field perturbations
on AdS Schwarzschild backgrounds were in focus at first, the holographic description
of the gravitational quasi-normal modes were also investigated and led to some important
developments. The calculations are based on the method of holographic renormalization,
\cite{mans}, \cite{skenderis1}, \cite{skenderis2}, \cite{kraus}, \cite{skenderis3},
which under the
appropriate boundary conditions at spatial infinity yields the energy-momentum
tensor of the gravitational background on the bulk; this method puts on firm ground
a previous proposal for the definition of quasi-local energy in gravitational
theories, \cite{york}, and overcomes its limitations. One of the most spectacular
results derived in this context in recent years has been the connection between
black holes and relativistic
hydrodynamics and, in particular, the derivation of a universal value for the
ratio of shear viscosity to entropy density, known as KSS bound, \cite{kss1},
\cite{kss2}. They complement quite nicely the old ideas on black hole hydrodynamics
that led to the membrane paradigm, \cite{price}.

$AdS_4/CFT_3$ correspondence is less studied in the literature up to this date and
some new features in the holographic description of four-dimensional gravity may arise.
As far as the previous discussion is concerned, the axial-polar duality
among the gravitational perturbations of $AdS_4$ black holes may have an interesting
manifestation in the three-dimensional field theory at the conformal boundary of
space-time. Hopefully, it may also help to explain in more fundamental terms why there
is an underlying supersymmetric quantum mechanics in the mathematical description
of the gravitational quasi-normal modes of four-dimensional black holes.
Here, we present some new results in this direction
and reformulate (at least part of) the problem as black hole energy-momentum
tensor/Cotton tensor duality using general boundary conditions on the wave-functions
of the effective Schr\"odinger problems for axial and polar perturbations. In this way,
the gravitational Chern-Simons action on the dual conformal boundary comes into play,
since its energy momentum tensor is by definition the Cotton tensor in three
dimensions. Further details and applications of the correspondence will be
presented elsewhere.

The main material of this paper is based on the theory of quasi-normal modes
and the method of holographic renormalization for computing the boundary energy-momentum
tensor. Section 2 contains an overview of the gravitational perturbations
of black holes in four space-time dimensions with emphasis on the $AdS_4$
Schwarzschild background. We will not include the results of numerical investigations
that have been carried out in detail and appear in several research and review
papers. We derive, however, the asymptotic expansion of the metric perturbations
at spatial infinity that will be useful in the calculations. Section 3 contains
an account of the holographic computation of the energy-momentum tensor in
four-dimensional linearized Einstein gravity and then proceeds with its
evaluation under
general boundary conditions on the wave-functions of the effective Schr\"odinger
equations. Some intermediate steps of the calculations described in sections
2 and 3 are given in Appendices A and B, respectively. Section 4 contains
some connections with the hydrodynamic representation of black hole
perturbations, while keeping the presentation superficially simple, and
selects a privileged set of boundary conditions by requiring that the shear
viscosity of axial and polar perturbations to be equal.
Section 5 contains as side remark the idea that the pure dissipative
hydrodynamic modes of black hole physics may be accounted by the normalized
Ricci flow (when suitably embedded into Einstein's equations with negative
cosmological constant) at the non-linear level.  Section 6 contains our main
observations on the boundary manifestation
of axial-polar duality based on the general formulae included in this paper.
It is also shown the this duality operates entirely within the KSS bound for the
ratio of shear viscosity to the entropy density of black holes, thus providing
a correspondence between black hole hydrodynamics and the gravitational
Chern-Simons theory.
Section 7 contains our conclusions and a small list of selected directions for
future work.

Throughout this paper, we set $ 8 \pi G = \kappa^2$ and Newton's constant
is normalized as $G = 1$ in the Schwarzschild metric. The
boundary conditions (Dirichlet or mixed) always refer to the wave-functions
of the effective Schr\"odinger problems at spatial infinity and not to the
metric perturbations themselves; of course one follows from the other.
We will also abuse the term ``supersymmetric quantum mechanics", since there
are no fermions here. The term ``spatial infinity" will always refer to $r = \infty$ 
in the radial direction of space-time. 
Finally, the perturbations of the metric (and related geometric quantities) are
complex for each quasi-normal mode. Apparently, real expressions will arise
by appropriate superposition, although it is not yet known (as far as we
can tell from the literature) whether these modes form a complete set in the
strick mathematical sense.

\section{Gravitational perturbations of $AdS_4$ black holes}
\setcounter{equation}{0}

In this section we review the basic features of linear perturbations
around the four dimensional Schwarzschild background,
\be
g_{\mu \nu} =
g_{\mu \nu}^{(0)} + \delta g_{\mu \nu} ~,
\ee
using the canonical decomposition
of $\delta g_{\mu \nu}$ into two distinct classes called axial and polar
perturbations, \cite{wheeler}. The resulting theory of quasi-normal modes is
formulated in the presence of cosmological constant and some special
features of $AdS_4$ black holes are discussed in detail (see also \cite{lemos},
\cite{moss}). Hopefully,
the present exposition can be of more general value to the interested
reader, as it contains a number of explicit results together with
the companion Appendix A.

\subsection{Generalities}

First, we recall for notational purposes some basic facts about black holes
that will be used throughout this paper.

Einstein equations in four space-time dimensions with cosmological constant
$\Lambda$,
\be
R_{\mu \nu} = \Lambda g_{\mu \nu} ~,
\ee
admit the Schwarzschild solution as spherically symmetric static
configuration of the form
\be
ds^2 = -f(r) dt^2 + {dr^2 \over f(r)} + r^2 \left(d\theta^2 +
{\rm sin}^2 \theta d\phi^2 \right)
\ee
with
\be
f(r) = 1 - {2m \over r} - {\Lambda \over 3} r^2
\ee
having the appropriate asymptotic behavior fixed by $\Lambda$.

The Schwarzschild radius of $AdS_4$ black holes is provided by the real root
of $f(r) = 0$ occurring at
\be
r_{\rm h} = {1 \over \sqrt{-\Lambda}} ~
\big[\left(\sqrt{1 - 9m^2 \Lambda}
+ 3m \sqrt{-\Lambda} ~ \right)^{1/3} -
\left(\sqrt{1 - 9m^2 \Lambda}
- 3m \sqrt{-\Lambda} ~ \right)^{1/3} \big] ~.
\ee
Thus, the black hole radius takes values $0 < r_{\rm h} < 2m$ depending
on the size of $\Lambda$. When $\Lambda$ approaches zero, $r_{\rm h}$ tends
to $2m$, whereas for $\Lambda << 0$, $r_{\rm h}$ comes close to 0.

It is also useful to introduce the tortoise coordinate $r_{\star}$ which
is defined by
\be
dr_{\star} = {dr \over f(r)} ~.
\ee
When $\Lambda = 0$, $r_{\star}$ ranges from $-\infty$ to $+\infty$,
as $r$ ranges from the black hole horizon located at $r=r_{\rm h}$ to
infinity. But when $\Lambda < 0$, which is of interest here, $r_{\star}$
ranges from $-\infty$ up to some finite value that can be set equal to
zero by appropriate choice of the integration constant. For $AdS_4$ black
holes, in particular, we have explicitly
\be
r_{\star}  =  {r_{\rm h} \over 4 (r_{\rm h} - 3m)}
\left(r_{\rm h} ~ {\rm log} {(2r + r_{\rm h})^2 + a^2  \over
4 (r-r_{\rm h})^2} +
2 a ~ {r_{\rm h} - 6m \over
r_{\rm h} + 6m} ~ \big[{\rm arctan}
{2r + r_{\rm h} \over a} -{\pi \over 2} \big]\right)
\ee
setting for convenience
\be
a = \sqrt{-{3 \over \Lambda} \left(1 + {6m \over r_{\rm h}} \right)} .
\ee

AdS black holes come in different sizes and their thermodynamic properties
depend crucially on the magnitude of $r_{\rm h}$ relative to the AdS radius
\be
L = \sqrt{-{3 \over \Lambda}} ~.
\ee
Large black holes have $r_{\rm h} > L$ and become the dominant configurations
at high temperatures, whereas small black holes have $r_{\rm h} < L$ and they
are always unstable to decay either into pure thermal radiation or to black
holes  with larger mass. In general we have the following relation among the
parameters of the $AdS_4$ Schwarzschild background
\be
m - r_{\rm h} = {1 \over 2L^2} r_{\rm h} \left(r_{\rm h}^2 - L^2 \right) .
\ee
Thus, large black holes have $r_{\rm h} < m$, whereas small black holes have
$r_{\rm h} > m$.

Finally, we recall that very large black holes are naturally
associated to the limit $r_{\rm h} \rightarrow \infty$, in which case
$f(r)$ is replaced by
\be
f(r) = -{2m \over r} - {\Lambda \over 3} r^2
\ee
by dropping the constant term. Then, the black holes become essentially flat
and their horizon is related to the other parameters by the simple
expression
\be
r_{\rm h}^3 = -{6m \over \Lambda} ~.
\ee

\subsection{Axial (odd) perturbations}

The first class of metric perturbations of four dimensional black holes is
tabulated by matrices labeled by $(t, r, \theta, \phi)$ of the following form
\be
\delta g_{\mu \nu} = \left(\begin{array}{cccc}
0 & 0 & 0 & h_0(r) \\
  &   &   &   \\
0 & 0 & 0 & h_1(r) \\
  &   &   &   \\
0 & 0 & 0 & 0 \\
  &   &   &   \\
h_0(r) & h_1(r) & 0 & 0
\end{array} \right)
e^{-i\omega t} {\rm sin} \theta ~
\partial_{\theta} P_l ({\rm cos} \theta) ~,
\ee
using the Legendre polynomials $P_l({\rm cos} \theta)$. More general
expressions in terms of spherical harmonics $Y_l^m (\theta, \phi)$
can also be employed, but one may only use axially symmetric
perturbations, setting $m=0$ without loss of generality. Axial perturbations
correspond to the so called vector sector or shear channel in the dictionary of
AdS/CFT correspondence.

The linear gravitational perturbations $\delta R_{\mu \nu} = \Lambda
\delta g_{\mu \nu}$ about the Schwarzschild background
yield the following equation for the $(\theta \phi)$-component,
\be
h_0(r) = i {f(r) \over \omega} \left(f(r) h_1(r)\right)^{\prime} ~,
\ee
whereas the equation for the $(r\phi)$-component reads
\be
{2 \over r} h_0(r) - h_{0}^{\prime}(r) = i {f(r) \over \omega}
\left({\omega^2 \over f(r)} - {(l-1)(l+2) \over r^2} \right)
h_1(r) ~.
\ee
These form a coupled system of first order differential
equations for the unknown functions $h_0(r)$ and $h_1(r)$, which are
otherwise unrelated.
The $(t\phi)$-component gives rise to a second order differential equation,
which, however, is trivially satisfied by virtue of the previous two
equations. All other components of $\delta R_{\mu \nu}$ are identically
zero and yield no further conditions.

Following Regge and Wheeler, \cite{wheeler}, we define the following variable
\be
\Psi_{\rm RW} (r) = {f(r) \over r} h_1(r) ~,
\ee
which turns out to satisfy the effective one-dimensional Schr\"odinger
equation
\be
\left(-{d^2 \over dr_{\star}^2} + V_{\rm RW}(r) \right) \Psi_{\rm RW}(r)
= \omega^2 \Psi_{\rm RW}(r)
\ee
with respect to the tortoise coordinate $r_{\star}$ with potential
\be
V_{\rm RW}(r) = f(r) \left({l(l+1) \over r^2} - {6m \over r^3} \right) .
\ee
Thus, one is led to consider solutions of the Regge-Wheeler-Schr\"odinger
problem by imposing appropriate boundary conditions (typically ingoing
at the black hole horizon and outgoing at spatial infinity), which in
turn can determine $\Psi_{\rm RW}(r)$
(and hence $h_1(r)$ and subsequently $h_0(r)$) together with the
allowed spectrum of quasi-normal mode frequencies $\omega$.

$V_{\rm RW}$ depend on $l$ and represent spherically symmetric
potentials surrounding the black hole. Plotting the potentials
as function of the tortoise radial coordinate can only be made
numerically because $r$ cannot be expressed in terms of $r_{\star}$
in closed form. For $\Lambda =0$ the potentials are manifestly positive
everywhere and extend on the real line $-\infty < r_{\star} < \infty$ falling
off to zero at both ends. For $\Lambda < 0$, on the other hand, the potentials
extend on the half-line $-\infty < r_{\star} \leq 0$, becoming zero on the horizon
and reaching a finite positive value at spatial infinity. In this case, however,
$V_{\rm RW}$ are not always everywhere positive definite, but they can become
negative for sufficiently low values of $l$, namely for large black holes with
\be
{m \over r_{\rm h}} > {l(l+1) \over 6} ~,
\ee
thus exhibiting a laguna. Although the differences between large and
small $AdS_4$ black holes leave their footprints on the shape of the
effective potential barriers for sufficiently small values of $l$,
their plots are alike for large values of $l$ exhibiting a maximum peak
followed by a local minimum as $r$ increases towards spatial infinity.

\subsection{Polar (even) perturbations}

This is a complementary class of metric perturbations parametrized by
four arbitrary radial functions of the general form
\be
\delta g_{\mu \nu} = \left(\begin{array}{cccc}
f(r)H_0(r) & H_1(r) & 0 & 0 \\
  &   &   &   \\
H_1(r) & H_2(r)/f(r) & 0 & 0 \\
  &   &   &   \\
0 & 0 & r^2K(r) & 0 \\
  &   &   &   \\
0 & 0 & 0 & r^2K(r) {\rm sin}^2 \theta
\end{array} \right)
e^{-i\omega t}
P_l ({\rm cos} \theta)
\ee
They correspond to the so called scalar sector or
sound channel in the dictionary of AdS/CFT correspondence.
Study of such linear
perturbations $\delta R_{\mu \nu} = \Lambda g_{\mu \nu}$ about the
Schwarzschild background yields
\be
H_0(r) = H_2(r) ~.
\ee
This choice will be made from the beginning to simplify the remaining
equations.

Tedious computation shows that the $(tr)$- $(r\theta)$- and
$(t\theta)$-components of the perturbation yield the following
equations, respectively,
\ba
& & r K^{\prime}(r) + \left(1 - {rf^{\prime}(r) \over 2f(r)}
\right) K(r) - H_0(r) - i {l(l+1) \over 2 \omega r} H_1(r) = 0 ~, \\
& & \left(f(r) H_0(r)\right)^{\prime} - f(r) K^{\prime}(r)
+ i\omega H_1(r) = 0 ~, \\
& & \left(f(r) H_1(r)\right)^{\prime} + i\omega \left(H_0(r)
+ K(r) \right) = 0 ~.
\ea
Together they form a coupled system of first order differential
equations for the three unknown functions $H_0(r)$, $H_1(r)$ and $K(r)$.
The other components of the perturbation either yield second order equations or
else $\delta R_{\mu \nu}$ vanishes identically.
Note here, however, that there is an additional algebraic condition among the
three radial functions
\ba
& & \left(2f(r) - rf^{\prime}(r) - l(l+1) \right) H_0(r) +
{i \over 2\omega} \left(4\omega^2 r - l(l+1) f^{\prime}(r) \right)
H_1(r) = \nonumber \\
& & \left(2f(r) + rf^{\prime}(r) - l(l+1) + 2\Lambda r^2 +
{r^2 \over 2f(r)} \left(4 \omega^2 + {f^{\prime}}^2(r) \right)
\right) K(r) ~,
\ea
which follows from consistency of the various second order equations with
the first order system above; it can also be viewed as integral of the
first order system above.

Following Zerilli, \cite{zerilli}, we now define the following variable
\be
\Psi_{\rm Z} (r) = {r^2 \over (l-1)(l+2)r + 6m} \left(K(r)
- i{f(r) \over \omega r} H_1 (r) \right) ,
\ee
which turns out to satisfy an effective Schr\"odinger equation, as
before,
\be
\left(-{d^2 \over dr_{\star}^2} + V_{\rm Z}(r) \right) \Psi_{\rm Z}(r)
= \omega^2 \Psi_{\rm Z}(r)
\ee
with different potential,
\ba
V_{\rm Z}(r) & = & {f(r) \over [(l-1)(l+2) r + 6m]^2}
\left(l(l+1)(l-1)^2 (l+2)^2 - 24m^2 \Lambda \right. \nonumber \\
& & \left. + {6m \over r} (l-1)^2 (l+2)^2
+ {36m^2 \over r^2}(l-1)(l+2)
+ {72 m^3 \over r^3} \right) .
\ea
Again, one has to find solutions and determine the quasi-normal mode
spectrum under appropriate boundary conditions, as before. This will,
in turn, lead to expressions for the three unknown radial functions
of polar perturbations.

As before, $V_{\rm Z}$ depend on $l$ and represent spherically symmetric
potential barriers surrounding the black hole, which are always positive
definite reaching a finite value at spatial infinity.
For $\Lambda < 0$, the shape of the potential depends on the
size of the black hole. In fact, $V_{\rm Z}$ appear to increase monotonically
for large black holes with sufficiently low values of $l$, whereas for
large values of $l$ they exhibit a maximum peak
followed by a local minimum as $r$ increases towards spatial infinity.
In these cases, $V_{\rm Z}$ resemble the shape of $V_{\rm RW}$, but they
rise higher than them for given $l$.

\subsection{Supersymmetric partner potentials}

The Regge-Wheeler and Zerilli potentials admit the following representation
\be
V_{\rm RW}(r) = W^2(r) - {d W(r) \over dr_{\star}} + \omega_{\rm s}^2
\ee
and
\be
V_{\rm Z}(r) = W^2(r) + {d W(r) \over dr_{\star}} + \omega_{\rm s}^2
\ee
in terms of a suitably chosen real (positive) function
\be
W(r) = {6m f(r) \over r[(l-1)(l+2)r + 6m]} + i \omega_{\rm s} ~,
\ee
setting
\be
\omega_{\rm s} = -{i \over 12m} (l-1)l(l+1)(l+2) ~.
\ee

Thus, the two Schr\"odinger problems under investigation take the closely
related form
\be
\left(-{d^2 \over dr_{\star}^2} + W^2 \mp {dW \over dr_{\star}}\right)
\Psi (r_{\star}) = (\omega^2 - \omega_{\rm s}^2) \Psi (r_{\star})
\ee
and resemble supersymmetric partner potentials generated by the
superpotential $W(r_{\star})$, \cite{chandra1}, \cite{chandra2}, \cite{lemos}.
The quantity $E= \omega^2 - \omega_{\rm s}^2$ serves as the
energy of the effective quantum mechanical problem, but unlike
conventional supersymmetric quantum mechanics, \cite{ed3}, \cite{susy},
it is not bounded below
by zero. In fact, due to the physical boundary conditions imposed on
the wave functions associated to perturbations of black holes, the
quantum theory is that of an open system and the energies (and hence
$\omega$) are in general complex.

Due to this relation, which is only present in four space-time dimensions,
the solutions are inter-connected using the conjugate pair of first order
operators
\be
A = {d \over dr_{\star}} + W (r_{\star}) ~, ~~~~~
A^{\dagger} = - {d \over dr_{\star}} + W (r_{\star}) ~.
\ee
The two effective Hamiltonians are simply written as
\be
H_{\rm RW} = A^{\dagger} A + \omega_{\rm s}^2 ~, ~~~~~
H_{\rm Z} = A A^{\dagger} + \omega_{\rm s}^2 ~.
\ee
Then, if $\Psi_{\rm RW} (r_{\star})$ is a solution of the Regge-Wheeler
equation with frequency $\omega$, the function
\be
A \Psi_{\rm RW} (r_{\star})
= i(\omega_{\rm s} - \omega) \Psi_{\rm Z} (r_{\star})
\label{roii1}
\ee
will be solution of the Zerilli equation with the same frequency.
Likewise, a solution of the Zerilli equation with frequency $\omega$
gives rise to solution of the Regge-Wheeler equation with
the same frequency, as
\be
A^{\dagger} \Psi_{\rm Z} (r_{\star})
= i(\omega_{\rm s} + \omega) \Psi_{\rm RW} (r_{\star}) ~.
\label{roii2}
\ee

These relations are particularly useful for justifying mixed
boundary conditions on the wave functions. For $\Lambda < 0$, one
typically imposes perfectly reflecting Dirichlet boundary conditions
on the wave functions at spatial infinity located at $r_{\star} =0$.
If the axial and polar perturbations satisfy simultaneously
\be
\Psi_{\rm RW} (r_{\star} = 0) = 0 = \Psi_{\rm Z} (r_{\star} = 0)
\ee
supersymmetric quantum mechanics will also imply the Neumann boundary
conditions
\be
{d \over d r_{\star}} \Psi_{\rm RW} (r_{\star}) \mid_{r_{\star} = 0}
= 0 = {d \over d r_{\star}} \Psi_{\rm Z} (r_{\star}) \mid_{r_{\star} = 0} ~,
\ee
which are too restrictive to hold all together. The conflict is resolved either by
abandoning supersymmetry, meaning that the spectrum of quasi-normal modes of axial
and polar perturbations is taken to be different, or by imposing mixed boundary
conditions as dictated by equations \eqn{roii1} and \eqn{roii2} above.

We also note for completeness that when $\Lambda = 0$ the boundary conditions
imposed at spatial infinity, namely outgoing waves for either axial or
polar perturbations, are compatible with supersymmetric quantum mechanics.

\subsection{Asymptotic expansions at spatial infinity}

The wave functions $\Psi_{\rm RW} (r)$ and $\Psi_{\rm Z} (r)$ represent incoming
waves to the black-hole. Therefore, they have the following
asymptotic expansion close to the horizon, following closely the analysis of
reference \cite{gary},
\be
\Psi (r)
= \sum_{n=0}^{\infty} a_n \left(1 - {r_{\rm h} \over r} \right)^n
~ e^{-i\omega r_{\star}} ~.
\ee
In either case, the coefficients $a_n$ depend upon $\omega$. They satisfy a
three-term recursion relation when substituted into the Regge-Wheeler equation
and a five-term recursion relation when substituted into the Zerilli equation.
Both power series expansions make good sense for all $r$ when $\Lambda < 0$,
because their radius of convergence extends to infinity. Then, the boundary
conditions at spatial infinity impose additional constraints on the coefficients
$a_n$, which in turn determine the spectrum of allowed quasi-normal modes of
$AdS_4$ black holes by numerical methods.

Here, we will reorganize the series expansion of the wave functions
in powers of $1/r$ to provide their asymptotic behavior at spatial infinity
for $\Lambda <0$.
The coefficients will be determined up to the order relevant for the
computation of the energy-momentum tensor for axial and polar perturbations of
$AdS_4$ black holes. Thus, these coefficients will be obtained under general boundary
conditions, but in the applications specific choices will be made at spatial infinity.

${\bf (i). ~ Axial ~ perturbations:}$ The asymptotic expansion of the Regge-Wheeler
wave function at spatial infinity is taken to be
\be
\Psi_{\rm RW} (r) = \left(I_0 + {I_1 \over r} + {I_2 \over r^2}
+ {I_3 \over r^3} + \cdots \right) e^{-i \omega r_{\star}} ~,
\ee
where the coefficients $I_k$ depend upon $\omega$ and they
are determined recursively from $I_0$ and $I_1$ as
\ba
{2 \Lambda \over 3} I_2 & = & 2i \omega I_1 - l(l+1) I_0 ~, \\
2 \Lambda I_3 & = & 4i \omega I_2 - (l-1)(l+2) I_1
+ 6m I_0
\ea
and so on. The boundary conditions at spatial infinity are solely expressed
in terms of $I_0$ and $I_1$.

The asymptotic expansion of the metric functions
$h_0 (r)$ and $h_1 (r)$ near spatial infinity are given in all
generality by
\ba
h_0 (r) & = & \left( \alpha_0 r^2
+ \beta_0 r + \gamma_0  + {\delta_0 \over r}
+ \cdots \right) e^{-i \omega r_{\star}} ~,  \\
h_1 (r) & = & \left({\alpha_1 \over r} + {\beta_1 \over r^2}
+ \cdots \right) e^{-i \omega r_{\star}} ~,
\ea
where the coefficients can be found in Appendix A after
expressing them in terms of $I_0$ and $I_1$ for convenience.

${\bf (ii). ~ Polar ~ perturbations:}$ Likewise, the asymptotic expansion of the
Zerilli wave function at spatial infinity is taken to be
\be
\Psi_{\rm Z} (r) = \left(J_0 + {J_1 \over r} + {J_2 \over r^2}
+ {J_3 \over r^3} + {J_4 \over r^4} + \cdots \right)
e^{-i \omega r_{\star}} ~,
\ee
where $J_k$ depend upon $\omega$ and they are determined recursively from
$J_0$ and $J_1$ via the relations
\ba
{2 \Lambda \over 3} J_2 & = & 2i \omega J_1 - \left(l(l+1) -
{24 m^2 \Lambda \over (l-1)^2 (l+2)^2} \right) J_0  ~, \\
2 \Lambda J_3 & = & 4i \omega J_2 - \left((l-1)(l+2) - {24 m^2 \Lambda
\over (l-1)^2 (l+2)^2} \right) J_1 + \nonumber\\
& & {6m \over (l-1)(l+2)} \left(l(l+1) + 2 - {48 m^2 \Lambda \over
(l-1)^2 (l+2)^2} \right) J_0 ~, \\
4 \Lambda J_4 & = & 6i \omega J_3 - \left(l(l+1) -6 - {24 m^2 \Lambda
\over (l-1)^2 (l+2)^2} \right) J_2 + \nonumber\\
& & {24m \over (l-1) (l+2)}
\left(1-{12 m^2 \Lambda \over (l-1)^2 (l+2)^2} \right) J_1 - \nonumber\\
& & {72 m^2 \over (l-1)^2 (l+2)^2} \left(l(l+1) + 1 -
{36 m^2 \Lambda \over (l-1)^2 (l+2)^2} \right) J_0
\ea
and so on. The boundary conditions at spatial infinity are solely expressed
in terms of $J_0$ and $J_1$, in analogy with the axial perturbations.

The asymptotic expansion of the metric functions $H_0 (r)$, $H_1(r)$ and
$K(r)$ take the following form at spatial infinity,
\ba
H_0 (r) & = & {3m \over (l-1)(l+2)}
\left({A_0 \over r} + {B_0 \over r^2} + {C_0 \over r^3}
+ \cdots \right) e^{-i \omega r_{\star}} ~, \\
H_1 (r) & = & -{3i \omega \over \Lambda}
\left({A_1 \over r} + {B_1 \over r^2} + {C_1 \over r^3} + \cdots \right)
e^{-i \omega r_{\star}} ~, \\
K(r) & = & \left(R + {A \over r} + {B \over r^2} + {C \over r^3}
+ \cdots \right) e^{-i \omega r_{\star}} ~.
\ea
The computations are much more
involved now and the various coefficients are
given explicitly in Appendix A expressing them in terms of
$J_0$ and $J_1$ alone.

These expansions are consistent with the differential equations as
well as with the algebraic constraint satisfied by the metric functions
of axial and polar perturbations with generic boundary conditions.
We also note for completeness that the wave functions could have been
expanded differently, e.g.,
\be
\Psi_{\rm RW} (r) =
\left(I_0^{\prime} + {I_1^{\prime} \over r} + {I_2^{\prime} \over r^2}
+ \cdots \right) e^{i \omega r_{\star}}  ~,
\ee
resembling the form of outgoing (rather than incoming) waves at
spatial infinity. The two expressions are equivalent provided that
$I_0^{\prime} = I_0$, $I_1^{\prime} = I_1 -(6i \omega / \Lambda) I_0$,
etc, using the asymptotic expansion $r_{\star}= 3/(\Lambda r) + \cdots$.
Similar remarks apply to the expansion of $\Psi_{\rm Z}(r)$ at
spatial infinity.

${\bf (iii). ~ Supersymmetric ~ partner ~ boundary ~ conditions:}$
If the spectrum of axial
and polar perturbations are related by supersymmetric quantum mechanics,
one should adopt mixed boundary conditions at spatial infinity that
are consistent with the general relation
\be
\left(- {d \over dr_{\star}} + W(r_{\star}) \right)
\Psi_{\rm Z} (r_{\star})
= i(\omega_{\rm s} + \omega) \Psi_{\rm RW} (r_{\star}) ~.
\ee
Then, the coefficients $J_0$ and $J_1$ arising in the asymptotic
expansion of $\Psi_{\rm Z}(r)$ should be related to the corresponding
coefficients $I_0$ and $I_1$ of $\Psi_{\rm RW}(r)$ as
\be
i(\omega_{\rm s} - \omega) J_0 = {\Lambda \over 3} I_1 +
\left(i(\omega_{\rm s} - \omega) - {2m \Lambda \over (l-1)(l+2)}
\right) I_0 ~,
\ee
\be
i(\omega_{\rm s} - \omega) J_1 = \left(i(\omega_{\rm s} +
\omega) - {2m \Lambda \over (l-1)(l+2)} \right) I_1 - \nonumber\\
\left(l(l+1) - {12 m^2 \Lambda \over (l-1)^2 (l+2)^2} \right)
I_0 ~,
\ee
Conversely, we also have
\be
i(\omega_{\rm s} + \omega) I_0 = - {\Lambda \over 3} J_1 +
\left(i(\omega_{\rm s} + \omega) - {2m \Lambda \over (l-1)(l+2)}
\right) J_0 ~,
\ee
\be
i(\omega_{\rm s} + \omega) I_1 = \left(i(\omega_{\rm s} -
\omega) - {2m \Lambda \over (l-1)(l+2)} \right) J_1 +
\left(l(l+1) - {12 m^2 \Lambda \over (l-1)^2 (l+2)^2} \right)
J_0 ~.
\ee

The simplest possibility of this kind is to impose Dirichlet boundary condition
on the axial perturbations and mixed boundary condition on the polar perturbations,
so that the coefficients are fixed by the relations
\be
I_0 = 0 ~, ~~~~~ I_1 = {3i \over \Lambda} (\omega_{\rm s}
- \omega) J_0 ~, ~~~~~ J_1 = {3 \over \Lambda}
\left(i(\omega_{\rm s} + \omega) - {2m \Lambda \over (l-1)(l+2)} \right) J_0 ~.
\ee
Other special choices of boundary conditions will be made later.

\subsection{The sign of $Im \omega$}

The perturbations diminish at late times provided that $Im \omega < 0$, in which
case
\be
\tau = -{1 \over Im \omega}
\ee
provides the characteristic time scale for return to equilibrium. Otherwise, the
perturbations grow large at spatial infinity and stability is at stake;
also, in such cases, the linear approximation can not
be reliably used to extract the late time behavior of the energy-momentum tensor on
the boundary.
The sign of $Im \omega$ can be shown to be negative when perfectly reflecting
boundary conditions are imposed at spatial infinity for the polar perturbations
and the same is true for the axial perturbations, at least in those cases that
the Regge-Wheeler potential does not form a laguna.
Different boundary conditions at spatial infinity may affect the sign of
$Im \omega$ for some modes in the spectrum, but there is no general proof that this
is indeed the case.

The standard analytic argument to address this question, \cite{gary} (but see also
\cite{lemos}), and which is extended here
to general boundary conditions, starts with the observation that the substitution
$\Psi(r) = u(r) {\rm exp}(-i \omega r_{\star})$ yields the differential equation
\be
{d \over dr} \left(f(r) {d u(r) \over dr}\right) - 2i \omega {du(r) \over dr}
- {V(r) \over f(r)} u(r) = 0
\ee
for either Regge-Wheeler or Zerilli potentials. Multiplying it with
the complex conjugate function $\bar{u} (r)$
and integrating over $r$, we obtain (after integrating by parts the first term)
the following relation,
\be
\int_{r_{\rm h}}^\infty dr \left(f(r) | {du(r) \over dr} |^2 +
2i \omega \bar{u}(r) {du(r) \over dr} + {V(r) \over f(r)}
| u(r) |^2 \right) = f(r) \bar{u}(r) {du(r) \over dr} \mid_{r=\infty} ~.
\ee
The right-hand side is simply $(\Lambda / 3) J_1 \bar{J}_0$ (respectively
$(\Lambda / 3) I_1 \bar{I}_0$) for polar (respectively axial) perturbations
satisfying general boundary conditions. Taking the imaginary part of this
integral equation, let us say for polar perturbations, and integrating
by parts the complex conjugate term, we obtain
\be
2i ~ Im \omega \int_{r_{\rm h}}^{\infty} dr \bar{u} (r) {du(r) \over dr} =
{\Lambda \over 3} Im (J_1 \bar{J}_0) - \bar{\omega} \left(|J_0|^2 -
|u(r_{\rm h})|^2 \right) .
\ee
This expression yields, upon substitution into the original
equation, the final result
\ba
\int_{r_{\rm h}}^\infty dr \left(f(r) | {du(r) \over dr} |^2 +
{V_{\rm Z} (r) \over f(r)} | u(r) |^2 \right) & = & {\Lambda \over 3}
J_1 \bar{J}_0 - {\Lambda \over 3} Im (J_1 \bar{J}_0) {\omega \over Im \omega}
+ \nonumber\\
& & {|\omega |^2 \over Im \omega} \left(| J_0 |^2 - | u(r_{\rm h}) |^2 \right) .
\ea
The left-hand side is positive definite and therefore $Im \omega < 0$ when
$J_0 = 0$. This argument is certainly inconclusive for more general boundary
conditions. For axial perturbations the coefficients are replaced by $I_0$ and
$I_1$ but the corresponding potential $V_{\rm RW}$ can become negative for large
black-holes when $l$ is sufficiently low. Thus, in those cases, the analytic
argument becomes inconclusive, even for $I_0 = 0$, but numerical analysis shows that
all such modes have $Im \omega < 0$.

In the following, we will assume that the spectrum of quasi-normal modes have
negative imaginary part either by selecting boundary conditions
for which this is manifest, by the argument above, or by employing numerical
methods to pin down those boundary conditions that yield
$Im \omega < 0$.

\section{Holographic energy-momentum tensor}
\setcounter{equation}{0}

In this section, we briefly review the construction of the boundary energy-momentum
tensor for AdS gravity, following \cite{kraus}, and apply it to the Schwarzschild
solution and its perturbations. The same results can be obtained using Fefferman-Graham
coordinates, as in the more systematic analysis of holographic renormalization presented
in references \cite{mans}, \cite{skenderis1}, \cite{skenderis2}, \cite{skenderis3}.
The companion Appendix B summarizes the results of intermediate steps in the
calculation.

\subsection{General considerations}

According to the AdS/CFT correspondence, the vacuum expectation
value of the energy-momentum tensor of the boundary quantum field
theory,
\be
<T^{ab}> = {2 \over \sqrt{-{\rm det} \gamma}} {\delta S_{\rm eff} \over
\delta \gamma_{ab}} ~,
\ee
is computed using the quasi-local energy-momentum tensor of a gravitational
bulk action $S_{\rm gr}$,
\be
T^{ab} = {2 \over \sqrt{-{\rm det} \gamma}} {\delta S_{\rm gr} \over
\delta \gamma_{ab}} ~.
\ee
The action $S_{\rm gr}$ which is defined on an asymptotically
AdS space-time $M$ is viewed as functional
of the boundary metric $\gamma_{ab}$ on $\partial M$. The resulting
$T^{ab}$ typically diverge, but it is always possible to obtain
finite results by adding an appropriately chosen boundary counter-term
whose form depends on the dimensionality of space-time. Holographic
renormalization provides a well defined prescription for implementing
the Brown-York procedure, \cite{york}, without using a reference space-time to
subtract the infinities.

In $AdS_4/CFT_3$ correspondence, in particular, the gravitational action
consists of bulk and boundary terms chosen as follows, \cite{kraus},
\ba
S_{\rm gr} & = & -{1 \over 2 \kappa^2} \int_{M} d^4 x \sqrt{-{\rm det} g}
\left(R[g] + 2 \Lambda \right) - {1 \over \kappa^2}
\int_{\partial M} d^3 x \sqrt{-{\rm det} \gamma} ~ K \nonumber\\
& & - {2 \over \kappa^2} \sqrt{-{\Lambda \over 3}} \int_{\partial M} d^3 x
\sqrt{-{\rm det} \gamma} \left(1 + {3 \over 4 \Lambda} R[\gamma]
\right) .
\ea
The first boundary contribution is the usual Gibbons-Hawking term written
in terms of the trace of the second fundamental form, i.e., the
extrinsic mean curvature
\be
K = \gamma^{ab} K_{ab} ~,
\ee
associated to the embedding of $\partial M$ in $M$. The second boundary
contribution is the contact term needed to remove all divergencies in
the present case.

Then, according to definition, the energy-momentum tensor of the
field theory is expressed in terms of the intrinsic and extrinsic
geometry of the AdS boundary at infinity, as
\be
\kappa^2 T_{ab} = K_{ab} - K \gamma_{ab} -2 \sqrt{-{\Lambda \over 3}}
\gamma_{ab} + \sqrt{-{3 \over \Lambda}} G_{ab} ~.
\ee
Here, $G_{ab}$ denotes the Einstein tensor of the induced
three-dimensional metric $\gamma_{ab}$,
\be
G_{ab} = R_{ab}[\gamma] - {1 \over 2} R[\gamma] \gamma_{ab} ~.
\ee
Clearly, only boundary terms contribute to the answer since the bulk
metric is always taken to satisfy the classical gravitational equations
of motion.

In practice, the computation is performed by first writing the metric
$g$ on $M$ in the form
\be
ds^2 = N^2 dr^2 + \gamma_{ab} \left(dx^a + N^a dr \right)
\left(dx^b + N^b dr \right)
\ee
using appropriately chosen $(N, N^a)$ functions, as in an ADM-like
decomposition. The three-dimensional surface arising at fixed distance
$r$ serves as boundary $\partial M_r$ to the interior
four-dimensional region $M_r$. The induced metric on $\partial M_r$
is $\gamma_{ab}$ evaluated at the boundary value of $r$, which is
held finite at this point. A useful relation among the bulk and boundary
metrics is
\be
\sqrt{-{\rm det} g} = N \sqrt{-{\rm det} \gamma} ~.
\ee

The second fundamental form $K_{ab}$ on $\partial M_r$ is defined
using the outward pointing normal vector $\eta_{\mu}$ to the
boundary $\partial M_r$ with components
\be
\eta_{\mu} = N ~ \delta_{\mu}^r ~.
\ee
In particular, one has
\be
K_{ab} = - \nabla_{(a} \eta_{b)} ~,
\ee
using the covariant derivatives with respect to the bulk metric $g$;
in the present case the expressions simplify to
\be
K_{ab} = N ~ \Gamma_{ab}^r [g] ~.
\ee
At the end of the computation, $T^{ab}$ on the AdS boundary $\partial M$
is obtained by letting $r \rightarrow \infty$.

Since the boundary metric acquires an infinite Weyl factor as $r$ is
taken to infinity, it is more appropriate to think of the AdS boundary
as a conformal class of boundaries and define $\mathscr{I}$ as the boundary
space-time with metric
\be
ds_{\mathscr{I}}^2 = \lim_{r \rightarrow \infty} \left(-{3 \over \Lambda r^2}
\gamma_{ab} dx^a dx^b \right) .
\ee
Then, the renormalized energy-momentum tensor on $\mathscr{I}$ is defined
accordingly by
\be
T_{ab}^{\rm renorm} = \lim_{r \rightarrow \infty} \left(\sqrt{-{\Lambda \over 3}}
~ r ~ T_{ab} \right)
\ee
and it is finite. This is the quantity that we will compute for all different type
of gravitational perturbations of $AdS_4$ black holes.

As for the trace of the energy-momentum
tensor on the three-dimensional boundary $\partial M_r$,
\be
\kappa^2 {T^a}_a = -2K -{1 \over 2} \sqrt{-{3 \over \Lambda}} R[\gamma]
- 6 \sqrt{-{\Lambda \over 3}} ~,
\ee
it has the following leading behavior for large $r$,
\be
{T^a}_a \sim {1 \over r^4} ~.
\ee
Terms of order $1/r^3$ are vanishing in this case by the absence of
conformal anomalies in three dimensions, \cite{mans}, and, therefore,
the trace of the renormalized energy-momentum tensor vanishes.

\subsection{Static $AdS_4$ black holes}

We first apply the formalism to the simple example of static $AdS_4$
Schwarzschild solution that will be subsequently used as reference
frame to study the effect of linear perturbations. All steps of the calculation
are included for illustrative reasons.

In this case, we have the following choice of $(N, N^a)$ functions,
\be
N={1 \over \sqrt{f(r)}} ~, ~~~~~ N^a = 0 ~,
\ee
and the induced metric on $\partial M_r$ is
\be
\gamma_{ab} = \left(\begin{array}{ccc}
-f(r) & 0  & 0 \\
      &    &  \\
0     & r^2   &  0\\
       &    & \\
0      & 0  & r^2 {\rm sin}^2 \theta
\end{array} \right) .
\ee

The second fundamental form turns out to be
\be
K_{ab} = \sqrt{f(r)} \left(\begin{array}{ccc}
f^{\prime}(r)/2 & 0  & 0 \\
      &    &  \\
0     & -r   &  0\\
       &    & \\
0      & 0  & -r {\rm sin}^2 \theta
\end{array} \right)
\ee
and its trace is
\be
K = -{1 \over 2r \sqrt{f(r)}} \left(r f^{\prime} (r) + 4 f(r) \right) ~.
\ee

Also, the Ricci curvature tensor of the induced metric $\gamma$
takes the simple form
\be
R_{ab}[\gamma]  = \left(\begin{array}{cccc}
0    &  & 0  & 0 \\
     &   &    &  \\
0    &   & 1   &  0\\
     &    &    & \\
0    &    & 0  & {\rm sin}^2 \theta
\end{array} \right)
\ee
and the Ricci scalar curvature is
\be
R[\gamma] = {2 \over r^2} ~.
\ee

Then, following the general prescription for computing the energy-momentum
tensor in AdS gravity,
we find the following expressions on $\partial M_r$,
\ba
\kappa^2 T_{tt} & = & {f(r) \over r^2} \left(\sqrt{-{3 \over \Lambda}}
+ 2r^2 \sqrt{-{\Lambda \over 3}} - 2r \sqrt{f(r)} \right) , \\
\kappa^2 T_{\theta \theta} & = & {r \over \sqrt{f(r)}} \left(f(r) +
{r \over 2} f^{\prime} (r) \right) - 2r^2 \sqrt{-{\Lambda \over 3}}
\ea
and $T_{\phi \phi} = {\rm sin}^2 \theta ~ T_{\theta \theta}$,
whereas all other components are zero.

As $r \rightarrow \infty$, $\partial M_r$ is pushed away to
spatial infinity and the energy-momentum tensor admits the
following asymptotic expansion
\ba
\kappa^2 T_{tt} & = & {2m \over r} \sqrt{-{\Lambda \over 3}}
+ {1 \over 4r^2} \sqrt{-{3 \over \Lambda}} +
{\cal O}\left({1 \over r^3} \right) , \\
\kappa^2 T_{\theta \theta} & = & {m \over r} \sqrt{-{3 \over \Lambda}}
+ {1 \over 4r^2} \left(\sqrt{-{3 \over \Lambda}}\right)^3 +
{\cal O}\left({1 \over r^3} \right) ,
\ea
whereas $T_{\phi \phi} = {\rm sin}^2 \theta ~
T_{\theta \theta}$, as before.
Note at this point that the trace of the energy-momentum
tensor is
\be
\kappa^2 {T^a}_a =
{1 \over 4r^4} \left(\sqrt{-{3 \over \Lambda}}\right)^3 +
{\cal O}\left({1 \over r^5} \right)
\ee
exhibiting the correct asymptotic behavior due to the absence of
conformal anomaly.

The three-dimensional metric on $\mathscr{I}$ is the Lorentzian
conformally flat metric on $R \times S^2$ written in spherical coordinates,
\be
ds_{\mathscr{I}}^2 = -dt^2 -{3 \over \Lambda}(d\theta^2 + {\rm sin}^2
\theta d\phi^2) ~.
\ee
The renormalized energy-momentum tensor of the boundary theory has
the following non-vanishing components
\be
\kappa^2 T_{tt}^{(0)} = -{2 m \Lambda \over 3} ~, ~~~~~
\kappa^2 T_{\theta \theta}^{(0)} = m ~, ~~~~~
\kappa^2 T_{\phi \phi}^{(0)} = m ~ {\rm sin}^2 \theta ~,
\ee
reproducing the expressions already known in the literature.
The superscript $(0)$ is used for reference to the static background.

\subsection{Axial perturbations}

Axial perturbations of AdS Schwarzschild black-holes are
parametrized by two radial functions $h_0(r)$ and $h_1(r)$.
The four-dimensional metric has coefficients
\be
N= {1 \over \sqrt{f(r)}} ~, ~~~~~
N_{\phi} = h_1(r) e^{-i \omega t} {\rm sin} \theta ~ \partial_{\theta}
P_l ({\rm cos} \theta) ~, ~~~~~ N_t = 0 = N_{\theta}
\ee
and the induced three-dimensional metric on $\partial M_r$ is
a perturbation of the static metric
\be
\gamma_{ab} = \gamma_{ab}^{(0)} +
\left(\begin{array}{ccc}
0  & 0  & h_0(r) \\
      &    &  \\
0     & 0   &  0\\
       &    & \\
h_0(r)      & 0  & 0
\end{array} \right) e^{-i\omega t} {\rm sin} \theta
~ \partial_{\theta} P_l ({\rm cos} \theta) ~.
\ee

The second fundamental form is also a perturbation of the
second fundamental form of the static solution,
\be
K_{ab} = K_{ab}^{(0)} + \delta K_{ab} ~
\ee
and the same thing applies to the Ricci curvature tensor of the
metric $\gamma_{ab}$,
\be
R_{ab}[\gamma] = R_{ab}^{(0)} + \delta R_{ab} ~.
\ee
However, the traces of $K_{ab}$ and $R_{ab}$ are inert to the
perturbations, i.e.,
\be
K=K^{(0)} , ~~~~~ R[\gamma] = R^{(0)} ~,
\ee
which in turn imply that the trace of the boundary energy-momentum
tensor coincides with the result obtained earlier for the static
background,
\be
{T^a}_a = {T^{(0)a}}_a ~.
\ee
This can be regarded as consistency check for the cancelation
of conformal anomalies for axial perturbations of the metric.

The complete energy-momentum tensor of the boundary theory on $\partial M_r$
assumes the following form,
\be
T_{ab}= T_{ab}^{(0)} +
\left(\begin{array}{ccc}
0  & 0  & \delta T_{t \phi} \\
      &    &  \\
0     & 0   &  \delta T_{\theta \phi} \\
       &    & \\
\delta T_{t \phi}  & \delta T_{\theta \phi}  & 0
\end{array} \right) ,
\ee
where $\delta T_{t \phi}$ and $\delta T_{\theta \phi}$ are given
explicitly in Appendix B.

Using the asymptotic expansion of the metric functions $h_0(r)$ and
$h_1(r)$ at spatial infinity, as given in Appendix A,
we find that all divergencies of $\delta T_{ab}$ cancel as
$r \rightarrow \infty$ irrespective of boundary conditions. In particular,
after conformal rescaling, the three-dimensional metric on $\mathscr{I}$
takes the form
\be
ds_{\mathscr{I}}^2 = -dt^2 -{3 \over \Lambda} \left(d \theta^2 +
{\rm sin}^2 \theta d \phi^2 \right) + 2 {iI_0 \over \omega}
e^{-i \omega t} {\rm sin} \theta ~ \partial_{\theta} P_l ({\rm cos} \theta)
~ dt d \phi
\ee
and the non-vanishing components of the axial perturbations of the
renormalized energy-momentum tensor are
\ba
\kappa^2 \delta T_{t \phi} & = & -{ i \Lambda \over 6 \omega}
\left(2m I_0 + (l-1)(l+2) \left({3i \omega \over \Lambda} I_0
-I_1 \right) \right) e^{-i \omega t} {\rm sin} \theta ~
\partial_{\theta} P_l({\rm cos} \theta) ~, \\
\kappa^2 \delta T_{\theta \phi} & = &
{1 \over 2} \left({3i \omega \over \Lambda}
I_0 - I_1 \right) e^{-i \omega t} {\rm sin} \theta [l(l+1)
~ P_l ({\rm cos} \theta) + 2 {\rm cot} \theta ~
\partial_{\theta} P_l ({\rm cos} \theta)] ~.
\ea
It can be verified independently, as consistency check, that the total
energy-momentum tensor is traceless and conserved on $\mathscr{I}$.

\subsection{Polar perturbations}

Polar perturbations of AdS Schwarzschild black-holes are parametrized by
three radial functions $H_0 (r)$, $H_1 (r)$ and $K(r)$. In this case, the
four-dimensional metric admits an ADM-like decomposition with coefficients
\be
N = {1 \over \sqrt{f(r)}} \left( 1 + {1 \over 2} H_0(r) e^{-i \omega t}
P_l ({\rm cos} \theta) \right)
\ee
and
\be
N_t = H_1(r) e^{-i \omega t} P_l({\rm cos} \theta) ~, ~~~~~ N_{\theta}
= 0 = N_{\phi} ~.
\ee
Also, the induced three-dimensional metric on $\partial M_r$ is a perturbation
of the static metric with diagonal form
\be
\gamma_{ab} = \gamma_{ab}^{(0)} +  \left(\begin{array}{ccc}
f(r)H_0(r) & 0  & 0 \\
      &    &  \\
0     & r^2K(r)   &  0\\
       &    & \\
0      & 0  & r^2 K(r) {\rm sin}^2 \theta
\end{array} \right) e^{-i \omega t} P_l ({\rm cos}\theta) ~.
\ee

The second fundamental form is a perturbation of the corresponding static
expression, as before, and the same thing applies to the Ricci curvature
tensor of the corresponding metric $\gamma_{ab}$. It turns out that
the trace of the second fundamental form for polar perturbations is
\ba
K & = & K^{(0)} +
\left({1 \over 4 \sqrt{f(r)}} \left(f^{\prime} (r) H_0(r)
+ 2 f(r) H_0^{\prime} (r) + 4i\omega H_1(r) \right) \right. \nonumber\\
& & \left. ~~~~~~ + {1 \over r}
\sqrt{f(r)} \left(H_0 (r) - r K^{\prime} (r) \right) \right) e^{-i \omega t}
P_l ({\rm cos} \theta)
\ea
and the Ricci curvature scalar is
\be
R[\gamma] = R^{(0)} - {1 \over r^2} \left(\left(2 {\omega^2 r^2
\over f(r)} - (l-1)(l+2) \right) K(r) + l(l+1) H_0 (r) \right)
e^{-i \omega t} P_l ({\rm cos} \theta) ~.
\ee
Note that $\delta R_{\phi \phi} \neq {\rm sin}^2 \theta ~
\delta R_{\theta \theta}$, which will in turn imply that
$\delta T_{\phi \phi} \neq {\rm sin}^2 \theta ~ \delta T_{\theta \theta}$
for the corresponding components of the energy-momentum tensor.
It follows that the trace of the boundary energy-momentum on $\partial M_r$ is
not inert to these perturbations, since
\ba
\kappa^2 {T^a}_a & = & \kappa^2 {T^{(0) a}}_a -
\left(\left({2 \over r} \sqrt{f(r)} +
{f^{\prime} (r) \over 2 \sqrt{f(r)}} -
{l(l+1) \over 2 r^2} \sqrt{-{3 \over \Lambda}} \right)
H_0 (r) \right. \nonumber\\
& & \left. + \sqrt{f(r)} ~ H_0^{\prime} (r)
-{1 \over 2r^2} \sqrt{-{3 \over \Lambda}}
\left(2{\omega^2 r^2 \over f(r)} - (l-1)(l+2) \right) K(r) \right. \nonumber\\
& & \left. - 2 \sqrt{f(r)} ~ K^{\prime} (r)
+ {2i \omega \over \sqrt{f(r)}}
H_1 (r) \right) e^{-i \omega t} P_l ({\rm cos} \theta) ~.
\ea
However, as we will see shortly, the additional terms are of order ${\cal O}(1/r^4)$
when $r \rightarrow \infty$, in agreement with the cancelation of conformal
anomalies.

The complete energy-momentum tensor of the boundary theory on $\partial M_r$
takes a form that is complementary to the corresponding expression for axial
perturbations, namely
\be
T_{ab}= T_{ab}^{(0)} +
\left(\begin{array}{ccc}
\delta T_{tt}  & \delta T_{t \theta}  & 0 \\
      &    &  \\
\delta T_{t \theta} & \delta T_{\theta \theta}   &  0 \\
       &    & \\
0  & 0  & \delta T_{\phi \phi}
\end{array} \right) ,
\ee
where the corresponding expressions are given explicitly in Appendix B.

Using the asymptotic expansion of the metric functions $H_0(r)$, $H_1(r)$ and
$K(r)$ at spatial infinity, as given in Appendix A,
we find that all divergencies cancel as
$r \rightarrow \infty$ irrespective of boundary conditions and all works well
as required on general grounds. In this case, the
three-dimensional metric on the boundary takes the following form, after
conformal rescaling,
\be
ds_{\mathscr{I}}^2 = -dt^2 -{3 \over \Lambda} [1+ R e^{-i \omega t}
P_l ({\rm cos} \theta)] (d\theta^2 + {\rm sin}^2 \theta d \phi^2 ) ~,
\ee
where $R = K(r=\infty)$ is the following function of $\omega$
\be
R = {\Lambda \over 3} J_1 - \left(i \omega - {2m \Lambda
\over (l-1)(l+2)} \right) J_0 ~.
\ee

Explicit calculation shows that the non-vanishing components of the
polar perturbations of the renormalized energy-momentum tensor are:
\ba
\kappa^2 \delta T_{tt} & = & m \Lambda (R -
i \omega_{\rm s} J_0)
e^{-i \omega t} P_l ({\rm cos} \theta) ~, \\
\kappa^2 \delta T_{\theta \theta} & = & {1 \over 4}
\left(4m {l(l+1) + 1 \over (l-1)(l+2)} R - l(l+1) \left(1 +
{3 \omega^2 \over \Lambda} \right) J_0 \right) e^{-i \omega t}
P_l({\rm cos} \theta) + \nonumber\\
& & {1 \over 4} \left({12m R \over (l-1)(l+2)} - \left(l(l+1)
+ {6 \omega^2 \over \Lambda} \right) J_0 \right)
e^{-i \omega t}
{\rm cot} \theta ~ \partial_{\theta} P_l({\rm cos} \theta) , \\
\kappa^2 \delta T_{\phi \phi} & = & - {1 \over 4}
\left(4m {2l(l+1) - 1 \over (l-1)(l+2)} R - l(l+1) \left(l(l+1) - 1 +
{3 \omega^2 \over \Lambda} \right) J_0 \right) \times \nonumber\\
& & e^{-i \omega t}
{\rm sin}^2 \theta P_l({\rm cos} \theta) - \nonumber\\
& & {1 \over 4} \left({12m R \over (l-1)(l+2)} - \left(l(l+1)
+ {6 \omega^2 \over \Lambda} \right) J_0 \right)
e^{-i \omega t} {\rm sin} \theta {\rm cos} \theta ~
\partial_{\theta} P_l({\rm cos} \theta) ~, \\
\kappa^2 \delta T_{t \theta} & = & {1 \over 4} i \omega
(l-1)(l+2) J_0 e^{-i \omega t} \partial_{\theta} P_l({\rm cos} \theta) ~.
\ea
It can be verified, as consistency check, that the complete energy-momentum
tensor is traceless and conserved on $\mathscr{I}$.

Note that the renormalized $\delta T_{tt}$ vanishes only when
$R=i \omega_{\rm s} J_0$. These are mixed boundary conditions for the
polar perturbations that are supersymmetric partner to perfectly
reflecting boundary conditions, $I_0 = 0$, for the axial perturbations.

\section{Hydrodynamic representation}
\setcounter{equation}{0}

The energy-momentum tensor associated to the static $AdS_4$ black hole
represents a perfect conformal fluid on the three-dimensional boundary with
metric $g_{ab}^{(0)}$, velocity vector $u_a =(-1, 0, 0)$ and energy density
\be
\kappa^2 \rho = -{2m \Lambda \over 3} ~.
\ee
Thus, it makes sense to compare the fluctuations of the energy-momentum
tensor for linear perturbations of black holes with the theory
of first order hydrodynamics. The comparison is only formal in the general
case, but the representation of the results for the energy-momentum
tensor in terms of fluid dynamics will be helpful in the sequel.
The true hydrodynamic modes of black hole physics will also be discussed
in this section.

\subsection{First order hydrodynamics}

Recall that the energy momentum tensor of a perfect relativistic fluid takes the
following form
\be
T_{ab} = \rho u_a u_b + p \Delta_{ab} ~,
\ee
where
\be
\Delta_{ab} = u_a u_b + g_{ab}
\ee
is given in terms of the unit velocity vector $u_a u^a = -1$ and the metric.
Conformal fluids have energy-momentum tensor with zero trace and therefore $\rho = 2p$
in three dimensions.

Deviations from the perfect fluid form are parametrized by adding appropriate viscosity
terms. Since the hydrodynamic velocity is ambiguous for non-equilibrium processes one
should make a (physically insignificant) choice. We will use the so called {\em energy frame},
meaning that $u_a$ is the unit time-like eigenvector of $T_{ab}$ defined as
\be
T_{ab} u^b = - \rho u_a
\ee
Then, the energy-momentum tensor of a general relativistic fluid admits the following
decomposition (see, for instance, the textbook \cite{landau}),
\be
T_{ab} = \rho u_a u_b + p \Delta_{ab} + \Pi_{ab} ~,
\ee
where $\rho$, $p$ are the corresponding energy density and pressure fields.
$\Pi^{ab}$ is a transverse tensor, $u_a \Pi^{ab} = 0$, that describes the viscous
part of the energy-momentum tensor of the fluid, and, in general, it admits an expansion
in the derivatives of $u^a$,
\be
\Pi_{ab} = \Pi_{ab}^{(1)} + \Pi_{ab}^{(2)} + \cdots ~.
\ee

First order hydrodynamics is concerned with the structure of $\Pi_{(1)}^{ab}$ and
is well studied. In this case, using the energy frame, we have, \cite{landau},
\be
\Pi_{(1)}^{ab} = - \eta \sigma^{ab} - \zeta \Delta^{ab} (\nabla_c u^c) ~,
\ee
where
\be
\sigma^{ab} = 2 \nabla^{<a} u^{b>}
\ee
expresses the symmetric, transverse and traceless part of $\Pi^{ab}$
up to first derivatives in $u^a$. Here, we use the notation (adapted to three-dimensional
fluids) of the bracketed second rank tensor
\be
A^{<ab>} = {1 \over 2} \left(\Delta^{ac} \Delta^{bd}
(A_{cd} + A_{dc}) - \Delta^{ab} \Delta^{cd}
A_{cd} \right) ,
\ee
which is transverse, $u_a A^{<ab>} = 0$, and traceless, $g_{ab} A^{<ab>} = 0$.
The coefficients $\eta$ and $\zeta$
depend in general on $\rho$ and they are called shear and bulk viscosity,
respectively. Of course, conformal fluids have $\zeta = 0$, whereas the value of
$\eta$ depends on the particular case.

In this context, one may also consider the vorticity of the velocity vector field
$u_a$, which is defined as follows,
\be
\Omega^{ab} = {1 \over 2} \Delta^{ac} \Delta^{bd} (\nabla_c u_d -
\nabla_d u_c) ~,
\ee
and it is clearly antisymmetric. As will be seen shortly, axial and polar
perturbations can be distinguished from each other by their vorticity tensor
field.

\subsection{Formal identifications}

Applying first order hydrodynamics to the perturbations of $AdS_4$ black holes
we arrive at the following formal identifications regarding the shear viscosity
coefficient:

${\bf (i). ~ Axial ~ perturbations:}$ Using the energy-momentum tensor computed
for axial perturbations with
general boundary conditions and the associated metric on $\mathscr{I}$, one
easily finds that the normalized time-like unit vector $u_a$ has
components
\be
u_t = -1 ~, ~~~~~ u_{\theta} = 0
\ee
and
\be
u_{\phi} =
-{i \over 6m \omega} (l-1)(l+2) \left({3i \omega \over \Lambda} I_0 - I_1
\right) e^{-i \omega t} {\rm sin} \theta ~ \partial_{\theta} P_l ({\rm cos}
\theta)
\ee
within the linear approximation. Also, the corresponding energy density is
\be
\kappa^2 \rho = -{2m \Lambda \over 3} ~,
\ee
as in the unperturbed black hole case.

Explicit computation shows that all components of $\Pi_{ab}^{(1)}$ vanish within
the linear approximation apart from
\be
\kappa^2 \Pi_{\theta \phi}^{(1)} = {1 \over 2} \left({3i \omega \over \Lambda} I_0 -
I_1 \right) e^{-i \omega t} {\rm sin} \theta [l(l+1) P_l ({\rm cos} \theta) +
2 {\rm cot} \theta ~ \partial_{\theta} P_l ({\rm cos} \theta)] ~.
\ee
Likewise, the computation of $\sigma_{ab}$ also shows that all its components vanish
apart from $\sigma_{\theta \phi}$. The result turns out to be identical to
$\Pi_{\theta \phi}$ up to an overall factor that determines the coefficient $\eta$
of shear viscosity for axial perturbation. Direct comparison, within the context of
first order hydrodynamics, yields
\be
\kappa^2 \eta = {3im \omega S \over (l-1)(l+2) (I_0 + S)} ~,
\ee
where it is set for convenience
\be
S = {(l-1)(l+2) \over 6m} \left({3i \omega \over \Lambda} I_0 -
I_1 \right) .
\ee

As special case, we refer to axial perturbations satisfying perfectly reflecting
boundary conditions, $I_0 = 0$, for which it turns out that
\be
\kappa^2 \eta = {3im \omega \over (l-1)(l+2)} ~.
\ee

The axial perturbations have a vorticity field with non-vanishing component
\be
\Omega^{\theta \phi} = - {\Lambda^2 \omega_{\rm s} \over 9 \omega}
\left({3i \omega \over \Lambda} I_0 - I_1\right) e^{-i \omega t} {\rm sin} \theta
~ P_l ({\rm cos} \theta)
\ee
under general boundary conditions.

${\bf (ii). ~ Polar ~ perturbations:}$ Similar considerations for polar perturbations satisfying
general boundary conditions yield the normalized time-like unit vector with components
\be
u_t = -1 ~, ~~~~~ u_{\phi} = 0 ~,
\ee
and
\be
u_{\theta} =
{i \omega \over 4 m \Lambda} (l-1)(l+2) J_0 e^{-i \omega t}
\partial_{\theta} P_l ({\rm cos} \theta) ~,
\ee
whereas the corresponding energy density turns out to be
\be
\kappa^2 \rho = -{2m \Lambda \over 3} + m \Lambda (R-i \omega_{\rm s} J_0)
e^{-i \omega t} P_l ({\rm cos} \theta)  ~.
\ee

Explicit computation of the tensor $\Pi_{ab}^{(1)}$  yields
\ba
\kappa^2 \Pi_{\theta \theta}^{(1)} &=& {1 \over 8} \left({12m R \over (l-1)(l+2)}
- \left(l(l+1) + {6 \omega^2 \over \Lambda} \right) J_0 \right)  e^{-i \omega t}
\times \nonumber\\
& & [l(l+1) P_l ({\rm cos} \theta) +
2 {\rm cot} \theta ~ \partial_{\theta} P_l ({\rm cos} \theta)]
\ea
and
\be
\Pi_{\phi \phi}^{(1)} = - {\rm sin}^2 \theta ~ \Pi_{\theta \theta}^{(1)} ~,
\ee
in agreement with its traceless property. All other components
of $\Pi_{ab}$ vanish within the linear approximation. To compare with first order
hydrodynamics we also compute $\sigma_{ab}$ and find that its components
vanish apart from $\sigma_{\theta \theta}$ and $\sigma_{\phi \phi}$. Their
expressions are proportional to $\Pi_{\theta \theta}$ and $\Pi_{\phi \phi}$,
respectively, and comparison yields the following coefficient $\eta$ of shear
viscosity for polar perturbations,
\be
\kappa^2 \eta = -{i m \Lambda \over 2 \omega (l-1)(l+2)}
\left({12m R \over (l-1)(l+2) J_0}
- \left(l(l+1) + {6 \omega^2 \over \Lambda} \right) \right) .
\ee

The special case of polar perturbations with mixed
boundary conditions $R = i \omega_{\rm s} J_0 $, which are supersymmetric partner
to perfectly reflecting boundary conditions on the axial perturbations, leads to
the coefficient
\be
\kappa^2 \eta = {3im \omega \over (l-1)(l+2)} ~.
\ee
This value is identical to the shear viscosity of axial perturbations with
perfectly reflecting boundary conditions.

The polar perturbations always have vanishing vorticity, which distinguishes them
from the axial perturbation.

\subsection{True hydrodynamic modes}

The hydrodynamic representation of the energy-momentum tensor of black hole
perturbations is just a convenient (yet formal) way to rewrite the results
of the calculation.
Nevertheless, there is a fundamental relation between the physics of black holes
and relativistic hydrodynamics that goes beyond first order and extends to
higher order causal theories of fluid dynamics, \cite{israel}, \cite{lind},
\cite{kss3}, \cite{huben}, \cite{makoto}. The hydrodynamic equations can be
thought as an effective theory describing the dynamics of the system at large length
and time scales. The true hydrodynamic modes of black hole perturbations are
identified by computing the retarded two-point Green functions of the
energy momentum tensor and finding their behavior at zero spatial momentum for low
frequencies (for an overview, see, for instance, \cite{kss2}, and references
therein). A rather general result has emerged in this context in recent years,
namely that the
ratio of shear viscosity to the entropy density of a very large AdS black hole
assumes a universal value, \cite{kss1}. More precisely, it turns out that the true
hydrodynamic modes have shear viscosity
\be
\kappa^2 \eta = {m \over r_{\rm h}}
\ee
that is independent of $l$, and, therefore, the ratio of shear viscosity to entropy
density is
\be
{\eta \over s} = {4 \over r_{\rm h}^2} \left(-{3 \over \Lambda} \right)
\eta = {1 \over 4 \pi}
\ee
in units where Boltzmann's constant and Planck's constant are set equal to 1.
This result appears to be valid in all dimensions and it has been further argued
that it provides an absolute lower bound (known as KSS bound) for the ratio
$\eta / s$ of all substances in nature;
see also the general presentations \cite{kss2} by the same authors.

The first example of true hydrodynamic modes is provided by purely dissipative
modes with frequencies
\be
\Omega_{\rm s} = - i {(l-1)(l+2) \over 3 r_{\rm h}} ~,
\label{riorio}
\ee
which turn out to belong to the spectrum of axial perturbations satisfying
Dirichlet boundary conditions, $I_0 = 0$, up to ${\cal O}(1/r_{\rm h}^2)$
corrections, \cite{moss}, \cite{pufu}. Thus, for very large $AdS_4$ black
holes the values $\Omega_{\rm s}$ are exact and the corresponding shear
viscosity coefficient, as calculated earlier, is
\be
\kappa^2 \eta = {3im \omega \over (l-1)(l+2)} = {m \over r_{\rm h}}
\label{matsin}
\ee
and yields the KSS value. Polar perturbations with mixed boundary conditions
$R = i \omega_{\rm s}$ also admit purely dissipative modes with frequencies
$\Omega_{\rm s}$ and yield the same result \eqn{matsin} for very
large black holes.

Another example of true hydrodynamic modes is provided by the complex values
of frequency
\be
\Omega_{\pm} = \pm \sqrt{-{\Lambda \over 6} l (l+1)} -
i {(l-1)(l+2) \over 6 r_{\rm h}} ~,
\label{riorio2}
\ee
which turn out to belong to the spectrum of polar perturbations satisfying
mixed boundary conditions $R=0$, up to ${\cal O}(1/r_{\rm h}^2)$ corrections,
\cite{pufu}. Thus, for very large $AdS_4$
black holes the values $\Omega_{\pm}$ are exact and the corresponding shear
viscosity coefficient, as calculated earlier, turns out to be
\be
\kappa^2 \eta = {i m \Lambda \over 2 \omega (l-1)(l+2)}
\left(l(l+1) + {6 \omega^2 \over \Lambda} \right) = {m \over r_{\rm h}} ~,
\label{matsin2}
\ee
up to ${\cal O}(1/r_{\rm h}^2)$ corrections, and it yields the KSS value, as
before. Axial perturbations with mixed boundary conditions
\be
{I_1 \over I_0} = {3 \over \Lambda} i \omega \left(1 - {\omega \over
\omega_{\rm s}} \right) + {6m \over (l-1)(l+2)}
\label{spebaou}
\ee
are supersymmetric partner to polar perturbations with $R=0$ and as such
they also admit quasi-normal modes with complex frequencies
$\Omega_{\pm}$. Comparison with the corresponding shear viscosity coefficient
yields the same result \eqn{matsin2} for very large black holes.

Actually, one can easily show that the only supersymmetric partner boundary
conditions that yield
\be
\eta_{\rm axial} = \eta_{\rm polar} ~,
\ee
as computed
explicitly in the previous subsection on general grounds, are
(i) $I_0 = 0$ and $R= i \omega_{\rm s} J_0$, and (ii) $I_1/I_0$ given by
equation \eqn{spebaou} and $R=0$; all other boundary conditions yield
$\eta_{\rm axial} \neq \eta_{\rm polar}$. Furthermore, by demanding
\be
\eta_{\rm axial} = \eta_{\rm polar} = m/r_{\rm h} ~,
\ee
it follows from the analysis above that the only allowed
frequencies are $\Omega_{\rm s}$ and $\Omega_{\pm}$ when $r_{\rm h}
\rightarrow \infty$.

Gravitational perturbations associated to true hydrodynamic modes
(of either type) satisfying the above special boundary conditions
will be particularly relevant in section 6.

\section{Connection with the normalized Ricci flow on $S^2$}
\setcounter{equation}{0}

The observation made in the literature, as result of numerical investigations,
that very large $AdS_4$ black holes exhibit purely dissipative modes for
axial perturbations satisfying perfectly reflecting Dirichlet boundary
conditions with frequencies \eqn{riorio}, $\Omega_{\rm s}$,
calls for an analytic explanation. The same set of modes also
arise for polar perturbations satisfying mixed boundary conditions that are
supersymmetric partner to the axial perturbations of very large $AdS_4$ black
holes with Dirichlet boundary conditions.

Recall at this point that there is a second order geometric evolution equation
for metrics on a Riemannian manifold driven by the Ricci curvature tensor,
\be
\partial_u g_{\mu \nu} = - R_{\mu \nu} ~,
\ee
known as Ricci flow (see, for instance, the collection of selected works
\cite{ricci}). The volume of space is not preserved under the evolution,
but it is always possible to define a variant, known as normalized Ricci flow,
which is volume preserving. The Ricci flow for the class of conformally flat metrics
on $S^2$,
\be
ds_2^2 = 2 e^{\Phi (z, \bar{z} ; u)} dz d \bar{z} ~,
\ee
takes the following form
\be
\partial_u \Phi = e^{-\Phi} \partial \bar{\partial} \Phi ~,
\ee
whereas the corresponding normalized Ricci flow on $S^2$ with fixed area
$4 \pi$ is given by
\be
\partial_u \Phi = e^{-\Phi} \partial \bar{\partial} \Phi + 1  ~.
\ee

The constant curvature metric provides the fixed point for the
normalized Ricci flow equation on $S^2$. In fact, the canonical
metric is reached from any given initial data after sufficiently long time.
It is instructive to examine the spectrum of linear perturbations around
this equilibrium state at late times,
using small axially symmetric deformations of the
round unit sphere parametrized by $\epsilon_l (u) P_l ({\rm cos} \theta)$,
\be
ds_2^2 = [1 + \epsilon_l (u) P_l ({\rm cos} \theta)]
\left(d\theta^2 + {\rm sin}^2 \theta d\phi^2 \right) ~.
\label{lorat}
\ee
It can be easily verified that the normalized Ricci
flow yields the following characteristic decay of metric perturbations,
as $u \rightarrow \infty$,
\be
\epsilon_l (u) = \epsilon_l (0) {\rm exp} \left(-{u \over 2}
(l-1)(l+2) \right) .
\ee
Then, the spectrum of purely imaginary frequencies associated to the
normalized Ricci flow is
given by
\be
\Omega_{\rm s} \sim - i {(l-1)(l+2) \over 2} ~,
\ee
up to a universal factor that depends on the physical scale of $u$ and
can be identified with $3r_{\rm h}/2$ to match the values \eqn{riorio}.

In view of this relation, it is natural to expect that there is an embedding
of the (normalized) Ricci flow into Einstein equations so that the
resulting four-dimensional metric describes a new radiative class of space-times.
In this context, $u$ should have the interpretation of retarded time
and the $AdS_4$ black hole should arise as a fixed point (static) configuration
after all radiation has been damped away. Also, in this context,
$\tau_{\rm s} = 1/i \Omega_{\rm s}$ should be the characteristic time scale,
depending on $l$, for the multi-pole gravitational radiation damping
close to equilibrium.
We do not expect this embedding to exist when $\Lambda = 0$
nor to be exact in the non-linear regime when the size of the black hole
is not very large. This idea might be more natural to implement in the polar
sector which resembles the perturbations \eqn{lorat} in the spherical
part of the four-dimensional metric.

It will also be interesting to have an analogous analytic explanation for the
existence of the complex frequencies $\Omega_{\pm}$ in the quasi-normal mode
spectrum of very large $AdS_4$ black holes. The boundary conditions are
different in this case and, therefore, the geometric framework that may account
for their presence will not be the same.

\section{Energy-momentum/Cotton tensor duality}
\setcounter{equation}{0}

In this section we describe the main application of the
results for the energy-momentum tensor of perturbed black holes.
We first introduce the notion of Cotton tensor in three dimensions, using
the Chern-Simons gravitational action, and then compare the two expressions
for suitably selected boundary conditions.

\subsection{Chern-Simons gravitational action}

In three dimensions there is a quantity that remains invariant under local conformal
changes of the metric $\gamma_{ab}$ and vanishes if and only if the metric is
conformally flat. It is provided by the density $\sqrt{{\rm det} \gamma }
~ {C^a}_b$, where $C_{ab}$ is an odd parity tensor, called Cotton tensor,
\ba
C^{ab} & = & {1 \over 2 \sqrt{- {\rm det} \gamma}} \left( \epsilon^{acd}
\nabla_c {R^b}_d + \epsilon^{bcd} \nabla_c {R^a}_d \right) \nonumber\\
& = & {\epsilon^{acd} \over \sqrt{- {\rm det} \gamma}} \nabla_c
\left({R^b}_d - {1 \over 4} {\delta^b}_d R \right)
\ea
with $\epsilon^{t \theta \phi} = 1$. The Cotton tensor is symmetric, traceless and
identically covariantly conserved. As such, it arises as functional derivative of a
geometric invariant, namely the three-dimensional gravitational Chern-Simons action,
\cite{cs},
\be
C_{ab} = {1 \over \sqrt{- {\rm det} \gamma}}
{\delta S_{\rm CS} \over \delta \gamma^{ab}} ~,
\ee
where
\be
S_{\rm CS} = {1 \over 2} \int d^3 x \sqrt{- {\rm det} \gamma} ~
\epsilon^{abc} \Gamma_{ae}^d \left(
\partial_b \Gamma_{cd}^e + {2 \over 3} \Gamma_{bf}^e \Gamma_{cd}^f \right) .
\ee
$S_{\rm CS}$ is an action of third order in the dynamical variables of
the theory.

The gravitational Chern-Simons action on the boundary of asymptotically locally
$AdS_4$ backgrounds arises from the topological Hirzebruch-Pontryagin
action on the bulk space-time, namely
\be
\int d^4 x \sqrt{- {\rm det} g} R_{abcd} {}^{\star} R^{abcd} =
{1 \over 2} \int d^4 x  \epsilon^{abef}  R_{abcd} {R_{ef}}^{cd} ~,
\ee
since the integrant is a total derivative; the $R \wedge R$ action enters into
the definition of the signature $\tau (M)$.
When the perturbations of black holes satisfy general boundary conditions, so
that the boundary metric is not conformally flat, the corresponding Cotton tensor
is non-vanishing. Thus, adding $S_{\rm CS}$ to the boundary action improves the
boundary energy-momentum tensor by the Cotton tensor and
changes the characteristics of the fluid velocity field in the hydrodynamic
representation of the problem; for example, the polar sector, which has no
vorticity, acquires some by this modification.
We will not pursue this general connection further in the present exposition.
Instead, we will restrict ourselves to the rather curious observation that the
Cotton tensor and the energy-momentum tensor of black hole perturbations exhibit
an {\em axial-polar duality} with respect to appropriately chosen supersymmetric
partner boundary conditions.

\subsection{The new correspondence for black holes}

We are now in position to establish the relation between the energy-momentum tensor
of black hole perturbations and the Cotton tensor of a dual boundary metric by
studying separately the polar and axial cases.

${\bf (i). ~ Polar ~ perturbations:}$ Let us first consider the boundary metric
for polar perturbations of $AdS_4$ black holes, which is given in general by
\be
ds_{\mathscr{I}}^2 ({\rm polar})= -dt^2 -{3 \over \Lambda} [1+ R e^{-i \omega t}
P_l ({\rm cos} \theta)] (d\theta^2 + {\rm sin}^2 \theta d \phi^2 ) ~.
\label{duba1}
\ee

Straightforward computation shows that its Cotton tensor has the following
non-vanishing components,
\ba
C_{\theta \phi} & = & {i \omega \over 4} R e^{-i \omega t} {\rm sin} \theta
[l(l+1) P_l ({\rm cos} \theta) + 2 {\rm cot} \theta ~ \partial_\theta P_l
({\rm cos} \theta)] ~, \\
C_{t \phi} & = & {\Lambda \over 12} (l-1) (l+2) R e^{-i \omega t}
{\rm sin} \theta ~ \partial_\theta P_l ({\rm cos} \theta) ~.
\ea
As such, they resemble the perturbations of the energy-momentum tensor for
axial perturbations. In fact, choosing the overall constant
\be
R= {2i \over \omega} I_1 ~,
\ee
the identification is exact provided that the energy-momentum tensor of
axial perturbations is evaluated at $I_0 = 0$, in which case the corresponding
boundary metric is conformally flat,
\be
ds_{\mathscr{I}}^2 ({\rm axial})= -dt^2 -{3 \over \Lambda}
(d\theta^2 + {\rm sin}^2 \theta d \phi^2 ) ~.
\label{duba2}
\ee

Thus, using the dual boundary metrics \eqn{duba1} and \eqn{duba2}, it follows
that
\be
C_{ab} ({\rm polar}) = \kappa^2 \delta T_{ab}({\rm axial})
\label{coreaa1}
\ee
for the supersymmetric partner boundary conditions
\be
R = i \omega_{\rm s} J_0 ~, ~~~~~  I_0 = 0
\ee
respectively, so that $\omega$ stays the same on both sides of the equality.

${\bf (ii). ~ Axial ~ perturbations:}$ Next, we consider the boundary metric
for axial perturbations of $AdS_4$ black holes, which is given in general by
\be
ds_{\mathscr{I}}^2 ({\rm axial})= -dt^2 -{3 \over \Lambda} \left(d \theta^2 +
{\rm sin}^2 \theta d \phi^2 \right) + 2 {iI_0 \over \omega}
e^{-i \omega t} {\rm sin} \theta ~ \partial_{\theta} P_l ({\rm cos} \theta)
~ dt d \phi ~.
\label{duba3}
\ee

In this case, the Cotton tensor of the metric has the following non-vanishing
components
\ba
C_{tt} & = & -{2m \Lambda^2 \over 3 \omega} \omega_{\rm s}  I_0
e^{-i \omega t} P_l ({\rm cos} \theta) ~, \\
C_{\theta \theta} & = & {i \Lambda \over 6 \omega} l(l+1) I_0
\left(1 + {3 \omega^2 \over \Lambda} \right) e^{-i \omega t} P_l
({\rm cos} \theta) + \nonumber\\
& & {i \Lambda \over 6 \omega} I_0 \left(l(l+1) + {6 \omega^2 \over \Lambda}
\right) e^{-i \omega t} {\rm cot} \theta ~ \partial_\theta P_l
({\rm cos} \theta) ~, \\
C_{\phi \phi} & = & - {i \Lambda \over 6 \omega} l(l+1) I_0
\left(l(l+1) - 1 + {3 \omega^2 \over \Lambda} \right) e^{-i \omega t}
{\rm sin}^2 \theta P_l ({\rm cos} \theta) \nonumber\\
& & - {i \Lambda \over 6 \omega} I_0 \left(l(l+1) + {6 \omega^2 \over \Lambda}
\right) e^{-i \omega t} {\rm sin} \theta  {\rm cos} \theta ~ \partial_\theta
P_l ({\rm cos} \theta) ~, \\
C_{t \theta} & = & {\Lambda \over 6} (l-1)(l+2) I_0 e^{-i \omega t}
\partial_\theta P_l ({\rm cos} \theta) ~,
\ea
which resemble the perturbations of the energy-momentum tensor for
polar perturbations. The identification becomes exact choosing
\be
I_0 = {3i \omega \over 2 \Lambda} J_0 ~,
\ee
provided that $R=0$, in which case the corresponding boundary metric is
conformally flat,
\be
ds_{\mathscr{I}}^2 ({\rm polar})= -dt^2 -{3 \over \Lambda}
(d\theta^2 + {\rm sin}^2 \theta d \phi^2 ) ~.
\label{duba4}
\ee

Thus, using the dual boundary metrics \eqn{duba3} and \eqn{duba4}, it follows
that
\be
C_{ab} ({\rm axial}) = \kappa^2 \delta T_{ab}({\rm polar})
\label{coreaa2}
\ee
for the supersymmetric partner boundary conditions
\be
R= 0 ~, ~~~~~ {I_1 \over I_0} = {3 \over \Lambda} i \omega \left(1 -
{\omega \over \omega_{\rm s}} \right) + {6m \over (l-1)(l+2)} ~,
\ee
respectively, so that $\omega$ is the same on both sides of the equality,
as before.
In this case, the perturbations satisfy mixed boundary conditions on both
sides of the relation.

Remarkably, the supersymmetric partner boundary conditions that realize
the energy-momentum/Cotton tensor duality for $AdS_4$ black holes are only
these ones with shear viscosity
\be
\eta_{\rm axial} = \eta_{\rm polar} ~.
\ee
Thus, the true hydrodynamic modes of very large black holes with frequencies
$\Omega_{\rm s}$ and $\Omega_{\pm}$, which fit precisely in this framework,
admit a new alternative description in terms of the three-dimensional
Chern-Simons gravitational action on the dual boundary. The perturbations of
the Schwarzschild metric at the conformal boundary, which arise on the
right-hand side of the correspondence \eqn{coreaa1} and \eqn{coreaa2}, are
simply zero,
\be
\delta g_{\mu \nu} \mid_{\mathscr{I}} = 0 ~.
\ee

\newpage

\section{Conclusions}
\setcounter{equation}{0}

We have computed the boundary energy-momentum tensor of $AdS_4$ black
holes for gravitational perturbations that satisfy arbitrary boundary
conditions at spatial infinity. The (yet mysterious) relation between
the effective Schr\"odinger problems for axial and polar perturbations,
which is best described by supersymmetric quantum mechanics, translates
into a duality between the energy-momentum and the Cotton tensor for
appropriately chosen boundary conditions at spatial infinity. This
framework accommodates the hydrodynamic modes of large $AdS_4$ black
holes, which satisfy the KSS bound $\eta / s = 1/ 4 \pi$, and, as such,
it can be viewed as a new correspondence operating on this bound.

Some related remarks have also appeared recently in the literature,
\cite{other1}, and in particular \cite{other2} that introduces the notion of  
dual gravitons on general grounds, but their manifestation in $AdS_4$
black hole backgrounds has not been made explicit. The results also
seem to be related to the (electric-magnetic) duality rotations of the
linearized four-dimensional Einstein equations, \cite{claudio} (but see 
also \cite{nieto} for earlier work), which
are formulated with no reference to Killing symmetries; for further
discussion and generalizations (including Einstein equations with
cosmological constant) we refer the reader to the literature
\cite{stan}, \cite{julia}, \cite{tassos}. Clearly, these connections
deserve further study that is left to future work. The applications
in $AdS_4/CFT_3$ correspondence at finite temperature in view of the
proposed correspondence with the gravitational Chern-Simons theory on
the dual boundary will also be investigated in detail in separate
publication.

Finally, another interesting question that emerged in this context is
the possibility to construct exact radiative metrics of vacuum
Einstein equations with negative cosmological constant, which settle
to large $AdS_4$ black holes and account for the special frequencies of
their hydrodynamic modes upon linearization. If this expectation
materializes, the hydrodynamic modes will be extended in the non-linear
regime and provide the gravity dual of non-linear hydrodynamics in
closed form. Embedding the Ricci flow into gravity seems to play a role
in this direction and it will also be investigated further in the future.

\vskip1.2cm

\centerline{\bf Acknowledgements}

This work was supported in part by the European Research and Training Network
``Constituents, Fundamental Forces and Symmetries of the Universe"
(MRTN-CT-2004-005104) and by the bilateral research grant
``Gravity, Gauge Theories and String Theory"
(06FR-020) for Greek-French scientific cooperation. I am grateful to the
organizers of the ``6th Spring School and Workshop on Quantum Field Theory
and Hamiltonian Systems" held in Calimanesti-Caciulata, Romania, 6-11 May 2008,
for providing the opportunity to present a first account of the main results
reported here.
Finally, I thank the theory group at CERN for hospitality and financial
support during the final stages of the present work; I have benefited by
attending the theory summer institute on black holes and from the exchanges I
had with other participants.

\newpage

\centerline{\bf Note added in v2} 

The bulk interpretation of the energy-momentum/Cotton tensor duality 
was investigated further by the author in the recent paper \cite{bakaioa}. 
There, it was found that spherical gravitational perturbations of $AdS_4$ space-time, 
which also split into axial and polar classes, are simply interchanged by 
the electric/magnetic duality of linearized gravity. In this simplified case, 
the axial and polar perturbations obey the same Schr\"odinger problem 
and thus the same boundary conditions at spatial infinity. The electric/magnetic 
duality of gravitational perturbations around $AdS_4$ space-time 
applies to all possible boundary conditions and it has holographic 
manifestation as energy-momentum/Cotton tensor duality at the conformal 
infinity. 

New features arise in the presence of black holes, since the axial and polar 
perturbations satisfy supersymmetric partner Schr\"odinger problems. Also,  
it is not known whether the electric/magnetic duality of linearized gravity in the 
bulk persists for perturbations around non-trivial backgrounds, such as the 
$AdS_4$ Schwarzschild solution. However, we believe that there is a remnant 
of duality in the linearlized theory, which explains the supersymmetric partnership 
of the black hole perturbations, althought it might not be applicable to all 
possible boundary conditions at spatial infinity. In fact, its validity might very well 
be restricted to the special boundary conditions singled out in the present work 
and provide the missing link for the bulk interpretation of the energy-momentum/Cotton 
tensor duality for $AdS_4$ black holes. In this context, the gravitational 
electric/magnetic duality will act as symmetry of the KSS bound, in analogy with 
$S$-duality of BPS states of gravitational theories; the same rational may also 
apply to the more general hydrodynamic relation $\eta_{\rm axial} = 
\eta_{\rm polar}$ under the previledged set of boundary conditions. 

These problems require separate investigation, which we intend to present 
elsewhere to illuminate their physical interpretation.

\newpage

\appendix
\section{Coefficients of the asymptotic expansion}
\setcounter{equation}{0}

In this appendix we summarize the first few coefficients in the asymptotic
expansion of the metric functions arising in the perturbations of
$AdS_4$ black holes. These are the only relevant terms for the computation
of the boundary energy-momentum tensor under general boundary conditions.

${\bf (i). ~ Axial ~ perturbations:}$ The coefficients of the metric function
$h_0(r)$ are
\ba
& & \alpha_0 = -{i \Lambda \over 3 \omega} I_0 ~, ~~~~
\beta_0 = I_0 ~, ~~~~
\gamma_0 = - i {(l-1)(l+2) \over 2 \omega} I_0 ~, \nonumber\\
& & \delta_0 = {l(l+1) \over 2 \Lambda} I_0 -{i \over 3 \omega}
\left((l-1)(l+2) + {3 \omega^2 \over \Lambda} \right) I_1 ~,
\ea
expressing them all in terms of $I_0$ and $I_1$ for convenience.
Likewise, the coefficients of the metric function $h_1(r)$ are
\be
\alpha_1 = -{3 \over \Lambda} I_0 ~, ~~~~~
\beta_1 = -{3 \over \Lambda} I_1 ~.
\ee

${\bf (ii). ~ Polar ~ perturbations:}$  The computations are much more
involved now and the expressions are quite cumbersome.
The coefficients in the asymptotic expansion of $H_0(r)$ turn out to be
\ba
A_0 & = & \left(
2i(\omega_{\rm s} + \omega) - {4 m \Lambda \over (l-1) (l+2)}
+ {\omega^2 \over m \Lambda}(l-1)(l+2) \right) J_0
-{2 \Lambda \over 3}  J_1 ~, \\
B_0 & = & (l-1)(l+2) \left(1 + {i \omega \over m \Lambda} \Big[1 -
{12 m^2 \Lambda \over (l-1)^2 (l+2)^2} \Big] \right) J_0 +
\nonumber\\
& & {(l-1)(l+2) \over 6m} \left((l-1)(l+2) + {6 \omega^2 \over
\Lambda} - {12i \omega m  \over (l-1)(l+2)}
\right) J_1 ~, \\
C_0 & = &
\left( - {(l-1)(l+2) \over 4m \Lambda} \Big[l(l+1) \left(l(l+1) - 4
\right) + {6 \omega^2 \over \Lambda} (l-1)(l+2) \Big] \right.
\nonumber\\
& & \left. + {6m \over (l-1)(l+2)} \Big[l(l+1) - 4 + {6 \omega^2 \over \Lambda}
\Big] + {6 i \omega \over \Lambda} \right) J_0 + \\
& & \left(l(l+1) - 4 + {6 \omega^2 \over \Lambda} +
{i \omega \over 2m \Lambda} (l-1)(l+2) \Big[(l-1)(l+2) +
{6 \omega^2 \over \Lambda} \Big] \right) J_1 ~.
\ea

Likewise, the coefficients in the asymptotic expansion of $H_1(r)$ are
given by
\ba
A_1 & = &
\left(i \omega - {2 m \Lambda \over
(l-1)(l+2)} \right) J_0
- {\Lambda \over 3} J_1 ~, \\
B_1 & = &
\left(l(l+1) - 1 - {12 m^2 \Lambda \over
(l-1)^2 (l+2)^2} \right) J_0
- \left(i \omega + {2 m \Lambda \over
(l-1)(l+2)} \right) J_1 ~, \\
C_1 & = &
3 \left(m {l(l+1) -4 \over (l-1)(l+2)} + {i \omega \over 2 \Lambda}
\Big[l(l+1) + 2 - {24 m^2 \Lambda \over (l-1)^2 (l+2)^2} \Big]
\right) J_0 + \nonumber\\
& & {1 \over 2} \left(l(l+1) - 4 - {6i \omega \over \Lambda}
\Big[i \omega + {2m \Lambda \over (l-1)(l+2)} \Big] \right)
J_1 ~.
\ea

Finally, the coefficients in the asymptotic expansion of $K(r)$ are
\ba
R & = & - A_1 ~, ~~~~~ B = {3i \omega \over \Lambda} A ~, \\
A & = &
- {1 \over 2} \left(l(l+1) - {24 m^2 \Lambda \over (l-1)^2 (l+2)^2}
\right) J_0 +
\left(i \omega + {2 m \Lambda \over (l-1) (l+2)} \right) J_1 ~, \\
C & = & - {1 \over 4 \Lambda} \left(l(l+1) \Big[l(l+1) - {12 \omega^2
\over \Lambda} - {24 m^2 \Lambda \over (l-1)^2 (l+2)^2} \Big]
\right. \nonumber\\
& & \left. ~~~~~~ + 12i \omega m  \Big[1 - {24 i \omega m \over
(l-1)^2 (l+2)^2} \Big] \right) J_0 \nonumber\\
& & + \left(m{l(l+1) \over (l-1)(l+2)} + {i \omega \over \Lambda}
\Big[1 - {6 \omega^2 \over \Lambda} + {12 i \omega m \over (l-1)(l+2)}
\Big] \right) J_1 ~.
\ea

In all expressions above the results are described entirely in terms of
the coefficients $J_0$ and $J_1$ for convenience, although this does not
particularly simplify the lengthy formulae.

\section{Energy-momentum tensor on $\partial M_r$}
\setcounter{equation}{0}

In this appendix we provide the intermediate results in the calculation
of the boundary energy-momentum tensor for perturbations of $AdS_4$
black holes.

${\bf (i). ~ Axial ~ perturbations:}$ Holographic renormalization yields
the following result for the perturbations of the energy-momentum
tensor on $\partial M_r$ in terms of the corresponding metric coefficients
$h_0(r)$ and $h_1(r)$,
\ba
\kappa^2 \delta T_{t \phi} & = & \left( \Big[{2 \over r} \sqrt{f(r)} +
{f^{\prime}(r) \over 2 \sqrt{f(r)}} - 2 \sqrt{-{\Lambda \over 3}} +
\sqrt{-{3 \over \Lambda}} {(l-1)(l+2) \over 2r^2} \Big]
h_0(r) \right. \nonumber\\
& & \left. - {1 \over 2} \sqrt{f(r)} \left(h_0^{\prime} (r) +
i \omega h_1 (r) \right) \right) e^{-i \omega t}
{\rm sin} \theta ~ \partial_{\theta} P_l ({\rm cos} \theta) ~, \\
\kappa^2 \delta T_{\theta \phi} & = & -{1 \over 2}
\left(\sqrt{f(r)} ~ h_1(r) + i \omega \sqrt{-{3 \over \Lambda}}
{h_0(r) \over f(r)} \right)
e^{-i\omega t} \times \nonumber\\
& & ~ {\rm sin} \theta [l(l+1) ~ P_l ({\rm cos} \theta) +
2{\rm cot} \theta  ~ \partial_{\theta} P_l ({\rm cos} \theta)] ~,
\ea
whereas the other components vanish.

${\bf (ii). ~ Polar ~ perturbations:}$ Likewise, we obtain
the following result for the perturbations
of the energy-momentum tensor on $\partial M_r$ in terms of the
corresponding metric functions $H_0(r)$, $H_1(r)$ and $K(r)$,
\ba
\kappa^2 \delta T_{tt} & = & f(r) \left(\Big[{3 \over r} \sqrt{f(r)} -
2 \sqrt{-{\Lambda \over 3}} - {1 \over r^2} \sqrt{-{3 \over \Lambda}}
~ \Big] H_0 (r) - \sqrt{f(r)} ~ K^{\prime} (r) \right. \nonumber\\
& & \left. ~~ + {(l-1)(l+2) \over 2r^2} \sqrt{-{3 \over \Lambda}} ~
K(r) \right) e^{-i \omega t} P_l ({\rm cos} \theta) ~, \\
\kappa^2 \delta T_{\theta \theta} & = & \left(\Big[r \sqrt{f(r)} +
{r^2 f^{\prime} (r) \over 2 \sqrt{f(r)}} - 2r^2 \sqrt{-{\Lambda \over 3}}
+ {\omega^2 r^2 \over 2f(r)} \sqrt{-{3 \over \Lambda}} ~ \Big]
K(r) \right. \nonumber\\
& & \left. ~~ + {r^2 \over 2} \sqrt{f(r)} ~ K^{\prime} (r) - i \omega
{r^2 \over \sqrt{f(r)}} H_1 (r) - {r^2 \over 2} \sqrt{f(r)} ~
H_0^{\prime} (r) \right. \nonumber\\
& & \left. ~~ -{r \over 2} \Big[\sqrt{f(r)} + {r f^{\prime} (r) \over
2 \sqrt{f(r)}} \Big] H_0 (r) \right) e^{-i \omega t}
P_l ({\rm cos} \theta) \nonumber\\
& & ~~ -{1 \over 2} \sqrt{-{3 \over \Lambda}} ~ H_0 (r) e^{-i \omega t}
{\rm cot} \theta ~ \partial_{\theta} P_l ({\rm cos} \theta) ~,  \\
\kappa^2 {\delta T_{\phi \phi} \over {\rm sin}^2 \theta} & = &
\left(\Big[r \sqrt{f(r)} +
{r^2 f^{\prime} (r) \over 2 \sqrt{f(r)}} - 2r^2 \sqrt{-{\Lambda \over 3}}
+ {\omega^2 r^2 \over 2f(r)} \sqrt{-{3 \over \Lambda}} ~ \Big]
K(r) \right. \nonumber\\
& & \left. ~~ + {r^2 \over 2} \sqrt{f(r)} ~ K^{\prime} (r) - i \omega
{r^2 \over \sqrt{f(r)}} H_1 (r) - {r^2 \over 2} \sqrt{f(r)} ~
H_0^{\prime} (r) \right. \nonumber\\
& & \left. ~~ -{r \over 2} \Big[\sqrt{f(r)} + {r f^{\prime} (r) \over
2 \sqrt{f(r)}} -{l(l+1) \over r} \sqrt{-{3 \over \Lambda}} ~ \Big]
H_0 (r) \right) e^{-i \omega t}
P_l ({\rm cos} \theta) \nonumber\\
& & ~~ +{1 \over 2} \sqrt{-{3 \over \Lambda}} ~ H_0 (r) e^{-i \omega t}
{\rm cot} \theta ~ \partial_{\theta} P_l ({\rm cos} \theta) ~, \\
\kappa^2 \delta T_{t \theta} & = & {1 \over 2} \left(i \omega
\sqrt{-{3 \over \Lambda}} ~ K(r) + \sqrt{f(r)} ~ H_1 (r) \right) e^{-i \omega t}
\partial_{\theta} P_l ({\rm cos} \theta) ~.
\ea


\newpage

\begin{thebibliography}{3}
\bibitem{wheeler}
T. Regge and J.A. Wheeler, ``Stability of a Schwarzschild singularity",
Phys. Rev. \underline{108} (1957) 1063.
\bibitem{zerilli}
F.J. Zerilli, ``Effective potential for even-parity Regge-Wheeler gravitational
perturbation equations", Phys. Rev. Lett. \underline{24} (1970) 737.
\bibitem{vish} C.V. Vishveshwara, ``Stability of the Schwarzschild metric",
Phys. Rev. \underline{D1} (1970) 2870;
L.A. Edelstein and C.V. Vishveshwara, ``Differential equations for
perturbations on the Schwarzschild metric", Phys. Rev.
\underline{D1} (1970) 3514.
\bibitem{chandra1}
S. Chandrasekhar and S. Detweiler, ``The quasi-normal modes of the
Schwarzschild black hole", Proc. Roy. Soc. Lond. \underline{A344}
(1975) 441.
\bibitem{chandra2}
S. Chandrasekhar, {\em The Mathematical Theory of Black Holes},
Oxford University Press, Oxford, 1983.
\bibitem{kokko}
K.D. Kokkotas and B.G. Schmidt, ``Quasinormal modes of stars and black holes",
Living Rev. Rel. \underline{2} (1999) 2 [gr-qc/9909058].
\bibitem{lemos}
V. Cardoso and J.P.S. Lemos, ``Quasinormal modes of Schwarzschild anti-de Sitter
black holes: Electromagnetic and gravitational perturbations", Phys. Rev.
\underline{D64} (2001) 084017 [gr-qc/0105103]; V. Cardoso, R. Konoplya and
J.P.S. Lemos, ``Quasinormal frequencies of Schwarzschild black holes in
anti-de Sitter space times: A complete study on the asymptotic behavior",
Phys. Rev. \underline{D68} (2003) 044024 [gr-qc/0305037].
\bibitem{moss}
I.G. Moss and J.P. Norman, ``Gravitational quasinormal modes for anti-de
Sitter black holes", Class. Quant. Grav. \underline{19} (2002) 2323
[gr-qc/0201016].
\bibitem{ed3} E. Witten, ``Dynamical breaking of supersymmetry", Nucl. Phys.
\underline{B188} (1981) 513; ``Constraints on supersymmetry breaking",
Nucl. Phys. \underline{202} (1982) 253.
\bibitem{susy}
F. Cooper, A. Khare and U. Sukhatme, ``Supersymmetry and quantum mechanics",
Phys. Rept. \underline{251} (1995) 267 [hep-th/9405029].
\bibitem{malda}
J. Maldacena, ``The large $N$ limit of superconformal
field theories and supergravity", Adv. Theor. Math. Phys.
\underline{2} (1998) 231 [hep-th/9711200].
\bibitem{igor}
S.S. Gubser, I.R. Klebanov and A.M. Polyakov, ``Gauge theory correlators
from noncritical string theory", Phys. Lett. \underline{B428} (1998) 105
[hep-th/9802109].
\bibitem{ed1}
E. Witten, ``Anti-de Sitter space and holography", Adv. Theor. Math. Phys.
\underline{2} (1998) 253 [hep-th/9802150].
\bibitem{ed2}
E. Witten, ``Anti-de Sitter space, thermal phase transition and confinement
in gauge theories", Adv. Theor. Math. Phys. \underline{2} (1988) 505
[hep-th/9803131].
\bibitem{don}
S.W. Hawking and D.N. Page, ``Thermodynamics of black holes in anti-de
Sitter space", Commun. Math. Phys. \underline{87} (1983) 577.
\bibitem{gary}
G.T. Horowitz and V.E. Hubeny, ``Quasinormal modes of AdS black holes and
the approach to thermal equilibrium", Phys. Rev. \underline{D62} (2000)
024027 [hep-th/9909056].
\bibitem{mans}
M. Henningson and K. Skenderis, ``The holographic Weyl anomaly", JHEP
\underline{9807} (1998) 023 [hep-th/9806087]; ``Holography and the Weyl
anomaly", Fortsch. Phys. \underline{48} (2000) 125 [hep-th/9812032].
\bibitem{skenderis1}
S. de Haro, S.N. Solodukhin and K. Skenderis, ``Holographic reconstruction
of space-time and renormalization in AdS/CFT correspondence", Commun. Math.
Phys. \underline{217} (2001) 595
[hep-th/0002230].
\bibitem{skenderis2}
K. Skenderis, ``Asymptotically anti-de Sitter space-times and their stress
energy tensor", Int. J. Mod. Phys. \underline{A16} (2001) 740 [hep-th/0010138];
``Lecture notes on holographic renormalization", Class. Quant. Grav.
\underline{19} (2002) 5849
[hep-th/0209067].
\bibitem{kraus}
V. Balasubramanian and P. Kraus, ``A stress tensor for anti-de Sitter gravity",
Commun. Math. Phys. \underline{208} (1999) 413 [hep-th/9902121].
\bibitem{skenderis3}
I. Papadimitriou and K. Skenderis, ``Thermodynamics of asymptotically locally
AdS spacetimes", JHEP \underline{0508} (2005) 004 [hep-th/0505190].
\bibitem{york}
J.D. Brown and J.W. York, ``Quasilocal energy and conserved charges
derived from the gravitational action", Phys. Rev. \underline{D47}
(1993) 1407.
\bibitem{kss1}
P.K. Kovtun, D.T. Son and A.O. Starinets, ``Viscosity in strongly interacting
quantum field theory from black hole physics", Phys. Rev. Lett.
\underline{94} (2005) 111601
[hep-th/0405231].
\bibitem{kss2}
P.K. Kovtun and A.O. Starinets, ``Quasi-normal modes and holography",
Phys. Rev. \underline{D72} (2005) 086009
[hep-th/0506184]; D.T. Son and A.O. Starinets, ``Viscosity, black holes and
quantum field theory", Ann. Rev. Nucl. Part. Sci. \underline{57} (2007) 95
[arXiv:0704.0240].
\bibitem{price}
R.H. Price and K.S. Thorne, ``Membrane viewpoint on black holes: Properties
and evolution of the stretched horizon", Phys. Rev. \underline{D33} (1986) 915;
K.S. Thorne, R.H. Price and D.A. Macdonald, {\em Black Holes: The Membrane
Paradigm}, Yale University Press, New Haven, 1986.
\bibitem{landau}
L.D. Landau and E.M. Lifshitz, {\em Fluid Mechanics}, second edition, Pergamon
Press, New York, 1987.
\bibitem{israel}
W. Israel, ``Non-stationary irreversible thermodynamics: A causal relativistic
theory", Ann. Phys. \underline{100} (1976) 310;
W. Israel and J.M. Stewart, ``Transient relativistic theormodynamics and
kinetic theory", Ann. Phys. \underline{118} (1979) 341.
\bibitem{lind}
W.A. Hiscock and L. Lindblom, ``Stability and causality in dissipative
relativistic fluids", Ann. Phys. \underline{151} (1983) 466.
\bibitem{kss3}
R. Baier, P. Romatschke, D.T. Son, A. Starinets and M.A. Stephanov,
``Relativistic viscous hydrodynamics, conformal invariance and holography",
JHEP \underline{0804} (2008) 100 [arXiv:0712.2451].
\bibitem{huben}
S. Bhattacharyya, V.E. Hubeny, S. Minwalla and M. Rangamani,
``Non-linear fluid dynamics from gravity",
JHEP \underline{0802} (2008) 045 [arXiv:0712.2456].
\bibitem{makoto}
M. Natsuume and T. Okamura, ``Causal hydrodynamics of gauge theory 
plasmas from AdS/CFT duality", Phys. Rev. \underline{D77} (2008) 
066014 [arXiv:0712.2916].
\bibitem{pufu}
G. Michalogiorgakis and S.S. Pufu, ``Low-lying gravitational modes in the
scalar sector of the global $AdS_4$ black holes", JHEP \underline{0702}
(2007) 023 [hep-th/0612065].
\bibitem{ricci}
H.-D. Cao, B. Chow, S.-C. Chu and S.-T. Yau, eds, {\em Collected Papers on
Ricci Flow}, Series in Geometry and Topology, vol. 37, International Press,
Somerville, 2003.
\bibitem{cs}
S. Deser, R. Jackiw and S. Templeton, ``Topologically massive gauge theories",
Ann. Phys. \underline{140} (1982) 372; Erratum-ibid. \underline{185}
(1988) 406; ``Three-dimensional massive gauge theories", Phys. Rev. Lett.
\underline{48} (1982) 975.
\bibitem{other1}
G. Compere and D. Marolf, ``Setting the boundary free in AdS/CFT", Class. Quant.
Grav. \underline{25} (2008) 195014 [arXiv:0805.1902].
\bibitem{other2}
S. de Haro, ``Dual gravitons in $AdS_4/CFT_3$ and the holographic Cotton tensor",
[arXiv:0808.2054].
\bibitem{claudio}
M. Henneaux and C. Teitelboim, ``Duality in linearized gravity", Phys. Rev.
\underline{D71} (2005) 024018 [gr-qc/0408101].
\bibitem{nieto}
J.A. Nieto, ``S-duality for linearized gravity", Phys. Lett. \underline{A262} 
(1999) 274 [hep-th/9910049]. 
\bibitem{stan}
S. Deser and D. Seminara, ``Duality invariance of all free bosonic and
fermionic gauge fields", Phys. Lett. \underline{B607} (2005) 317
[hep-th/0411169]; ``Free spin 2 duality invariance cannot be extended
to GR", Phys. Rev. \underline{D71} (2005) 081502 [hep-th/0503030].
\bibitem{julia}
B. Julia, J. Levie and S. Ray, ``Gravitational duality near de Sitter space",
JHEP \underline{0511} (2005) 025 [hep-th/0507262].
\bibitem{tassos}
R.G. Leigh and A.C. Petkou, ``Gravitational duality transformations on
$(A)dS_4$", JHEP \underline{0711} (2007) 079 [arXiv:0704.0531].
\bibitem{bakaioa}
I. Bakas, ``Duality in linearized gravity and holography", [arXiv:0812.0152]. 
\end{thebibliography}
\end{document}